# Dynein dynamics at the microtubule plus-ends and cortex during division in the *C. elegans* zygote

Ruddi Rodriguez-Garcia[1,2,3,*], Laurent Chesneau[1,2,*], Sylvain Pastezeur[1,2], Julien Roul[1,2,4], Marc Tramier[1,2], & Jacques Pécréaux[1,2,#]

[1] CNRS, UMR6290, Institute of Genetics and Development of Rennes (IGDR), 35043 Rennes, France. [2] University of Rennes 1, UEB, SFR Biosit, School of Medicine, 35043 Rennes, France. [3] Present address: Cell Biology, Faculty of Science, Utrecht University, Padualaan 8, 3584 CH Utrecht, The Netherlands. [4] Present address: LAAS - Laboratoire d'analyse et d'architecture des systèmes, 7 avenue du Colonel Roche, BP 54200, 31031 Toulouse cedex 4, France.

* These authors contributed equally to this work.

# Correspondence and requests for materials should be addressed to J.P. (email: jacques.pecreaux@univ-rennes1.fr)

Running title: Dynamics of dynein in the *C. elegans* zygote

Abstract

During asymmetric cell division, dynein generates forces, which position the spindle to reflect polarity and ensure correct daughter cell fates. The transient cortical localization of dynein raises the question of its targeting. We found that it accumulates at the microtubule plus-ends like in budding yeast, indirectly hitch-hiking on EBP-2$^{EB1}$ likely via dynactin. Importantly, this mechanism, which modestly accounts for cortical forces, does not transport dynein, which displays the same binding/unbinding dynamics as EBP-2$^{EB1}$.
At the cortex, dynein tracks can be classified as having either directed or diffusive-like motion. Diffusive-like tracks reveal force-generating dyneins. Their densities are higher on the posterior tip of the embryos, where GPR-1/2$^{LGN}$ concentrate, but their durations are symmetric. Since dynein flows to the cortex are non-polarized, we suggest that this posterior enrichment increases dynein binding, thus accounts for the force imbalance reflecting polarity, and supplements the regulation of mitotic progression via the non-polarized detachment rate.

Successful symmetric and asymmetric cell division relies on the precise positioning and orientation of the mitotic spindle, thus ensuring the correct partitioning of chromosomes and cell organelles. The microtubule molecular motor dynein is key to producing the causative forces[1-11], from yeast to humans. Dynein is localized at the cell cortex, and generates pulling forces on the astral microtubules that radiate from the spindle poles[12]. Cytoplasmic dynein (hereafter referred to simply as "dynein") is minus-end directed motor, going towards the spindle poles. Dynein proteins are involved in numerous functions. These include the spindle assembly checkpoint and spindle positioning during mitosis, but also retrograde vesicular traffic in interphase cells and neuronal axons[13]. Dynein is a dimer of a multi-subunit complex carrying out various functions that are specified by the choice of the subunits[14]. Well-conserved throughout evolution, dynein subunits include: heavy chain (HC); intermediate chain (IC); light intermediate chain (LIC); and light chain (LC)[15]. The HCs, which are members of the AAA ATPase protein superfamily are the force-producing components[16]. In vertebrates, various ICs ensure cargo binding specificity[14]. In contrast, in *Caenorhabditis elegans*, only one IC homolog exists, DYCI-1, and its depletion creates phenotypes which mostly reflect the loss of the HC[17,18]. Dynein is essential to spindle positioning in higher eukaryotes and needs to localize at the cell periphery[19,20]. Recently it was proposed that this cortical localization could be highly dynamic[20,21], raising the question of what mechanism dynein might use to efficiently target the cortex. Such a mechanism has only been solved in fungi, mainly budding yeast[8,22-29] and fission yeast[30]. In this latter, dynein diffuses along the microtubule lattice to reach the cortex[30]. In budding yeast, dynein is targeted to the cortex via a three-part mechanism. First, dynein is transported along astral microtubules which point towards the bud, by the kinesin-related protein Kip2p and the microtubule-associated protein Bik1p/CLIP170[25,27,28] (whose closest homolog is CLIP-1 in *C. elegans*). It then accumulates at the microtubule plus-end. Secondly, thought to be the predominant mechanism, dynein can come directly from the cytoplasm to accumulate at the plus-end[28]. In these two cases, Bik1p/CLIP170 is essential, as it tracks plus-ends directly or via Bim1p/EB1, making this latter superfluous[25,29,31]. But thirdly, dynein may also reach the cortex without requiring microtubules, thus independently of Bik1p/CLIP170[24,27,28]. It is thought, however, that dynein arriving at the cortex by this technique fails to anchor[8]. In contrast, in the other mechanisms, once dynein reaches the bud it is offloaded and anchored at the cortex by Num1p in a dynactin-dependent manner[23]. Dynein accumulation at the microtubule plus-ends is central to positioning the spindle across the bud, in *Saccharomyces cerevisiae*. Similar plus-end gathering have been found in many organisms, including mammalian cells[32,33]. It was observed for example *in vivo* in HeLa cells[32] and in neuronal progenitors in mice[34], and is generally assumed to have similar causes to those of fungi, although the details remain unclear. Beyond the molecular details of dynein accumulation, it is not clear whether such a gathering at the plus-end contributes to *transporting* dynein to the cortex, since EB1 proteins are accumulated but not transported[35]. This is an important issue considering that this dynein, once at the cortex, generates in cortical force essential to metazoan cell division[19].

Once at the cortex, dynein is involved in generating forces in response to polarity cues, pulling on astral microtubules to position and orient the mitotic spindle[19,20,36,37]. Although the role of these cortical forces is essential in a broad range of organisms[19], the mechanistic link between polarity and dynein-generated forces remains elusive. Beyond the obvious possibility of an asymmetry in the amount of dynein present at the cortex, it could also be caused by differential regulation or dynamics (a higher binding rate or lower unbinding rate on the stronger force side). Any of these possible mechanisms could account for an asymmetric distribution of *active* force generators, ending in a force imbalance. This is seen in the *C. elegans* zygote[38-40], where in response to polarity cues[39], a limited number of cortical dynein-driven[41] active force generators[40,42] pull on astral microtubules, causing forces that displace the spindle out-of-cell-centre. The same forces contribute to elongation and anaphase spindle oscillation, which provides an accurate readout of dynein activity at the cortex[40,41,43]. In this way, dynein belongs to the so-called "force-generating complex," which also includes LIN-5[NuMA], GPR-1[LGN] and GPR-2[LGN] (hereafter referred to as GPR-1/2 as these are 96% identical)[44,45]. These protein regulators can connect the complex to the cell membrane through Gα proteins[41] and can limit the number of active force generators[40,42,46,47]. LIS-1 is also required for most dynein functions during zygotic division[41,48], although it was suggested to have a regulatory role of microtubule plus-end accumulation in mammals[49-51]. In budding yeast, Pac1p/LIS1 is required for targeting dynein[29], inhibiting it before it is anchored[52]. Based on fine analysis of oscillation frequencies, we previously assumed that cortical force generators pull for a very short time, 1 s or less[21]. This makes sense considering the highly dynamic localization discussed recently[20], which points to a force imbalance caused by an asymmetry in dynamics rather than in dynein counts. Altogether, this turns the spotlight back onto the mechanism that translates polarity into a force imbalance, having consequences on cell fates and/or the balance between proliferation and differentiation[19,20,36,37,43,53,54].

To address this important question, we first focussed on the mechanisms for bringing dynein to the cortex, since these may contribute to asymmetry and cortical force imbalance. In particular, we wondered whether dynein is actively transported, *id est* moved towards the cortex by consuming an energy source. To decipher the dynamics in the cytoplasm and at the microtubule plus-ends *in vivo* during the initial mitosis of the *C. elegans* zygote, we combined advanced image processing and fluorescence correlation spectroscopy (FCS). Similarly, we investigated cortical dynamics to discover how the force asymmetry is encoded, directly observing dynein dynamics there to increase our quantitative understanding of the mechanics of cell division.

# Results

### DYCI-1::mCherry is a *bona fide* reporter of dynein dynamics.

To decipher the mechanism of dynein targeting to the cortex and its residency there, we investigated dynein *in vivo* dynamics by using a strain carrying a randomly integrated transgene that codes for the fluorescent mCherry-labelled dynein intermediate chain DYCI-1, expressed under its own promoter[55,56]. This strain phenocopied the N2 control strain. We began by checking whether our construct was functional. Indeed, the transgene rescued the *dyci-1(tm4732)* null allele, and no phenotype was visible at the one-cell stage when randomly integrated into a homozygous *dyci-1(tm4732)* background (Supplementary Text 1.1-2). We concluded that despite possible altered expression levels, DYCI-1::mCherry can perform native DYCI-1 functions.

We used spinning disk microscopy to image the strain TH163, carrying randomly integrated DYCI-1::mCherry. We investigated the dynamics both at the cortex and in the plane between the cover slip and the spindle, the lower spindle plane (LSP) (Fig. 1d). We revealed the motion of fluorescent spots by computing the standard deviation map (SDM, or temporal variance image), which represents the variation in intensity of each pixel over time[57] (see Materials and Methods). In the LSP, we observed both spindle and central spindle staining, with spots moving radially towards the cortex during metaphase and anaphase (Fig. 1a,b, S1a,c, and Movies S1-2). This is consistent with the spindle and dotty cytoplasmic localizations previously revealed by antibody staining of dynein heavy chain DHC-1[41,58] or by CRISPR/Cas9-assisted labelling of the same subunit[59]. At the cell cortex, we observed transient spots (Fig. 1c, S1e-g, and Movie S3). The spots visible in the LSP and at the cortex cannot be caused by over-expression, because even when endogenous *dyci-1* was suppressed through the *tm473* homozygous null allele, they were seen when only two copies of the transgene were integrated by MosSCI[60,61] (Supplementary Text 1.3, Fig. S1b-d, f-h, and Movies S4-5). These spots must therefore be physiological. Their motion suggests that dynein is recruited at the microtubule lattice or at the plus-ends, as happens with yeast[28]. However, before investigating this we validated our strain and developed a biophysics approach for going beyond the localization to address whether the underlying mechanism transports dynein and how it may contribute to cortical force generation.

We next tested whether these spots are dynamics. We analysed their dynamics at the cortex and observed a rapid turnover, around a second, which indicate that spots form and disappear dynamically (Movie S3, Fig. 2d), as expected[20,21]. Studies in budding yeast and mammalian cells have suggested that dynein can accumulate at the microtubule plus-ends. We observed spot lifetime in the LSP, comparable to the time it takes for a microtubule plus-end to cross the focal plane (Movie S1 and S2). To gain certainty, we compared the assembly kinetics of the spots with that of EB proteins at the microtubule plus-ends[62]. We used a strain labelled with both DYCI-1::mCherry and EBP-2::GFP, measured spot intensity in each channel by FCS (Fig. 3c) and found that both proteins shared a similar spot attachment kinetics at the microtubule plus-ends (Fig S2c and Supplementary Text 1.5). Fluorescence correlation spectroscopy allows for the monitoring of diffusion into and out of a small volume. Importantly, we checked that both proteins were not associated in the cytoplasm by Fluorescence cross-correlation spectroscopy (FCCS, Fig. S2b). FCCS monitors the co-diffusion of two labelled molecules entering and exiting from the focal volume. We concluded that the DYCI-1::mCherry spots were dynamic in both the LSP and at the cortex, and they are therefore biologically relevant.

We then decided to ensure that when endogenous DYCI-1 is present, DYCI-1::mCherry is still involved in cortical pulling force generation. We wondered whether labelled DYCI-1 colocalises with the other members of the cortical force generating complex, GPR-1/2 and LIN-5[41]. We crossed strains carrying randomly integrated DYCI-1::mCherry and GPR-1::YFP and viewed both dyes (Fig. S3g). We found that 30% of the cortical spots in DYCI-1::mCherry ($N = 8$ embryos) colocalised with a GPR-1::YFP ones. Because the cortical anchors are in limited number[21,46,47,63], all dynein cortical spots are not contributing to pulling forces (see last results section), thus a larger proportion of colocalization was not expected. We also reasoned that an abundance of GPR-1::YFP spots might cause artefactual colocalizations. To exclude this, we compared the GPR-1::YFP colocalizing with dynein spots and with a simulated distribution of the same number of randomly localised spots (Fig. S3g, Supplementary Text 1.6). We found a very significant difference[64], indicating that colocalization was not artefactual. We performed similar controls for all further colocalizations. We concluded that this dynein subunit is probably involved in cortical force generation. In a broader take on the same question, Schmidt and colleagues observed similar dynein spots using a mCherry::DHC-1 construct which colocalises with eGFP::LIN-5[59]. We went on to look for a functional indication that our labelled dynein is involved in cortical pulling force generation. To do so, we used our previously published "tube" assay[42], which reports the localisation of force generation events by creating cytoplasmic membrane invaginations. These are rare in normal condition (Supplementary Text 1.6) but upon weakening the actin myosin cortex, with a partial *nmy-2(RNAi)* to preserve polarity[42], these tubes are more numerous. The invagination distribution reflects the force imbalance, while the depletion of the cortical force generator complex or related proteins by RNAi significantly decreases their number[42]. We used this assay to assess whether our DYCI-1::mCherry construct was functional, able to generate cortical forces. We crossed DYCI-1::mCherry and PH::GFP membrane labelling strains (Fig. S3a) and viewed the cortex upon partial *nmy-2(RNAi)* (Movie S6, Fig. S3c-e). We tracked the DYCI-1::mCherry spots, and observed that half of the invaginations colocalised with the resulting tracks (Fig. S3f, Supplementary Text 1.6). We figured that in the other half of the cases, this was not seen because of detection limits imposed by high DYCI-1::mCherry cytoplasmic background fluorescence. Indeed, the threshold for detecting a spot over the background fluorescence was estimated to be 26 ± 4 dyneins per spot (Supplementary Text 1.7). Dynein spots typically appeared 0.4 s before invagination (Fig. S3c-e and Movie S7), suggesting that dynein-related pulling forces must create the invaginations. Overall, DYCI-1::mCherry colocalization with cortical force generating complex member GPR-1 at the cortex and presence on invaginations indicate that labelled dynein DYCI-1::mCherry is involved in force generation.

We next wondered whether DYCI-1 — and more particularly DYCI-1::mCherry — is generally associated with the dynein complex in its various functions. Indeed, even very partial depletion of DYCI-1 by RNA interference produces a phenotype similar to *dli-1(RNAi)*[65] (Fig. S4a). Furthermore, DYCI-1 is the only dynein intermediate chain in *C. elegans*, and is already known to associate with two dynein complex subunits, DLC-1 and DYRB-1[66,67]. Globally, this suggests that DYCI-1 works with the other dynein complex subunits. The next question was whether the labelled DYCI-1 fraction performs similarly, being generally present in dynein complex. When we partially depleted the light intermediate chain DLI-1, essential to dynein functioning in the zygote[65], we lost the DYCI-1::mCherry spots entirely ($N = 5$ embryos, Fig. S5n). We also did not see pronuclei meeting in 10/12 embryos, which indicates the penetrance of that treatment. Because dynein is gathered at the microtubule plus-end from the cytoplasm, we asked whether dynein complex was assembled and included DYCI-1::mCherry in this compartment. To test this, we did *in vivo* measurements of the DYCI-1::mCherry diffusion coefficient ($D$), which reflects the size of the labelled object. Using FCS, we obtained $D = 2.6 ± 0.7$ μm$^2$/s ($N = 9$ embryos, 38 spots), which is about 5 times less than that of PAR-6::mCherry (which has about 40% fewer amino acids than DYCI-1). This suggests that DYCI-1::mCherry is embedded into a large complex in the cytoplasm. Its diffusion coefficient corresponds to the value computed from *in vitro* experiments for the entire dynein dimer[68] (Supplementary Text 1.8). We concluded that overall, DYCI-1::mCherry reveals dynein localization during one-cell embryo division.

As part of our attempt to decipher the dynein targeting mechanism and how it contributes supplying the molecular motors to the cortical pool to generate pulling forces, we next investigated how many dyneins molecules are present in a single spot in the LSP. We estimated this using a spot association kinetics equation (Supplementary Text 1.4) to overcome weak spot intensities that preclude direct measurement. In a strain carrying the randomly integrated DYCI-1::mCherry and EBP-2::GFP constructs, we found 66 ± 5 dyneins and 185 ± 85 EBP-2 per spot ($N = 8$ embryos, 38 spots). In the strain carrying two DYCI-1::mCherry copies on top of endogenous DYCI-1, there were 29 ± 6 dyneins per spot ($N$ = 5 embryos, 66 spots). Because this estimate was indirect, we sought a secondary and independent approach. After subtracting the background, we measured the intensity of dynein spots and used the PAR-6::mCherry strain's background fraction to calibrate the intensity versus the number of particles in FCS volume (Supplementary Text 2.1 and Fig. S6)[69]. In the strain carrying randomly integrated DYCI-1::mCherry, we obtained 50 ± 13 particles per spot ($N$ = 6 embryos, 20 spots), which is consistent with our previous estimate. Overall, our data show that the randomly integrated DYCI-1::mCherry transgene, herein referred to simply as DYCI-1::mCherry, is a *bona fide* dynein reporter, instrumental for investigating its dynamics both in the cytoplasm and at the cortex.

### Dynein spots displayed a directed (flow-like) motion towards periphery in the cytoplasm.

We noticed above dynein spots moving towards the cell periphery (Fig. 1a,b and Movie S1-2), and wondered what caused this and what its contribution to cortical pulling force regulation might be. In particular, among the various possible molecular mechanisms presented above, we wondered whether it involved the microtubule lattice or only its plus-end. Further, we wanted to know if it *actively* transports the dynein (*id est* consuming ATP or more broadly energy to do so) or passively accumulates it, by auto-organization. We therefore designed a pipeline to analyse the motion of DYCI-1::mCherry spots in the LSP. First we denoised the images using the CANDLE algorithm[70] (Fig. S7a,b) and enhanced the spots using the Laplacian of Gaussian (LoG) spot-enhancer filter[71] (Fig. S7c and Supplementary Text 3.1). We then tracked the images with u-track[72] (Supplementary Text 3.2). We classified the tracks according to their motion: anisotropic, which we termed "directed" and which corresponds to transport or 1D diffusion; and isotropic, or "diffusive-like," disregarding whether the underlying mechanism is normal or anomalous diffusion[73]. We divided the directed tracks between those moving towards the cortex and those moving towards the centrosomes. We then applied a Bayesian classification approach (BCA, Supplementary Text 3.3)[74] to those directed tracks moving towards the cortex. We distinguished between *diffusive motion* (i.e. normal diffusion which can even be a one-dimensional random walk and is not limited, confined, or enhanced by a motor); *flow* (transport-like mechanisms, excluding 1D diffusion); and a mixture of both. To challenge this analysis pipeline, we used fabricated microscopy images having signal-to-noise ratios (SNR) that were similar to the experimentally observed ones (Fig. S8a-e and Supplementary Text 3.4, Movie S8-9). In this numerical experiment, we successfully recovered the particle localizations (Fig. S8h); the trajectory speeds and durations (Fig. S8fg); and the separations between directed and diffusive-like trajectories (Fig. S8i). This validated our analysis pipeline. We analysed dynein movies acquired in the LSP, finding mostly directed tracks (Fig. 2a-e and S9). Because tracks with diffusive-like motion have shorter durations, we reasoned that they might be misclassified as diffusive-like regardless of their real motions. In support of this, analysis of the simulated data revealed that whatever their real motion, short tracks tended to be classified as diffusive-like. We also found a few tracks that move from the cortex to the centrosome in a directed motion. Because they are rare ( 5 ± 3 %) and not expected to contribute to bringing dynein at the cortex to generate pulling force, we did not investigate these any further. BCA used on tracks directed towards the periphery revealed that the motion was likely to be in a flow (Fig. 2f). This is compatible with either the active transport of dynein along the microtubule lattice, or a dynein accumulation at the microtubule plus-ends, for instance after hitching a ride with an EB1 homolog. In this latter mechanism, dynein would use EB1 accessory proteins to track microtubule plus-ends[75], binding to it likely indirectly through dynactin as suggested by *in vitro* experiments[76].

## Dynein accumulates at microtubule plus-ends, but is not *transported* towards the periphery.

Studies in budding yeast and other organisms suggest that dynein could accumulate at microtubules plus-ends reminding EB1. Indeed, DYCI-1::mCherry spots in LSP colocalized with the EB1 homolog EBP-2::GFP (Fig. 3a,b and Movie S10) in about the same proportion as with the microtubule (Fig. S10b, Movie S11, and Supplementary Text 2.2). This suggests that dynein and EBP-2 share a common position at the growing microtubule plus-ends.

Since EBP-2 is not transported to the cortex by microtubule plus-ends but rather accumulates there[35], the dynamics of dynein accumulation should be investigated. We found similar microtubule plus-end binding kinetics for EBP-2::GFP and DYCI-1::mCherry. In both cases the kinetics depend exponentially on their neighbouring cytoplasmic concentrations (Fig. S2c)[62]. This suggests that like EB1, dynein is mostly recruited from the cytoplasm. Interestingly, when examining these two proteins in the cytoplasm by FCCS, we found that they are not bound in the cytoplasm (Fig. S2b). This indicates that dynein has in fact its own plus-end attachment dynamics. Furthermore, this binding could be indirect. This again raised the tantalizing question of dynein transportation to the cortex. Indeed, using FCS we measured the exponential decay of dynein intensity along the microtubule lattice from the plus-end (comet-tail, Fig. 3c), and we obtained similar detachment dynamics for both DYCI-1::mCherry and EBP-2::GFP (Fig. 3d). We thus decided to test whether a dynein complex is just briefly localized in the microtubule plus-end, and whether it could display a detachment dynamics similar to the one of EB1 proteins[35]. In this case, we would expect the dynein spots to form a comet with an even longer tail, as the microtubule grows faster. We measured the DYCI-1::mCherry spot comet-tails (typically in 7 embryos, 1500 tracks per condition) when microtubule dynamics were modulated through hypomorphic *klp-7(RNAi)* or *clip-1(RNAi)* (Supplementary Text 2.3 and Fig. S2a). We found a linear relation between the comet-tail lengths and growth rates, with a slope significantly different from zero: 1.2 ± 0.2 s ($p = 0.03$) (Fig. 3e). Importantly, penetrant depletion of these genes did not preclude dynein recruitment at the plus-ends (Fig. 4b, S5l). Therefore, due to the similar binding/unbinding dynamics, we concluded that like EBP-2$^{EB1}$, dynein accumulates at the microtubule plus-ends rather than being actively transported by them.

## Dynein accumulates at microtubule plus-ends with the help of EBP-2, Dynactin and LIS-1, but independently from EBP-1/-3, CLIP-1$^{CLIP170}$, and probably also kinesins.

Because DYCI-1::mCherry displayed similar dynamics to EBP-2::GFP and colocalized with it, we asked whether dynein could track microtubule plus-ends using EBP-2$^{EB1}$. Indeed, *in vitro* experiments using purified human proteins[76] have led to the hypothesis that in higher eukaryotes dynein tracks the microtubule plus-ends via a hierarchical interaction involving dynactin, which in turn binds to EB1 with the help of CLIP170. This contrasts with previous findings in budding yeast, indicating that dynein is recruited at the microtubule plus-ends independently of dynactin[9,24,77] while Bik1p/CLIP-1[8,9,24-29,31] makes Bim1p/EB1 superfluous. Because of this differences, we wondered which of these mechanisms is at work in the nematode. We depleted EBP-2 by RNAi and observed severely decreased dynein densities in the directed (and diffusive) tracks of the LSP (Fig. 4a and S5a). However, no alteration was observed in the diffusion coefficient in the cytoplasm (Fig. S5b), indicating that dynein is not sequestered somewhere else. We confirmed this result by crossing the DYCI-1::mCherry strain with one carrying the *ebp-2(gk756)* null mutation. In this way we were able to obtain a viable strain without dynein spots in the cytoplasm (Fig. 4a), although some faint spots that occur below our detection limits may remain. To further test whether dynein hitches a ride with EBP-2, either directly or with the help of accessory proteins, we depleted dynactin and LIS-1 by *dnc-1(RNAi)* and *lis-1(RNAi)*, respectively. These RNAi were partial to preserve the early steps of mitosis. We observed a strong reduction in the LSP directed track densities (Fig. 4b and S5c-e), and as expected the treatment resulted in a strong phenotype reminiscent of dynein depletion[48,78]. In contrast, EBP-2::GFP plus-end accumulation was not affected by these same treatments (Fig. S5j), which is consistent with the previous observation that microtubule growth rates remain unaltered[79]. Furthermore, Schmidt and colleagues have reported that dynein plus-end accumulations colocalize with the DNC-1[p150glued] and DNC-2[p50/dynamitin] dynactin subunits[59]. Because these results are reminiscent of what was proposed in higher eukaryotes, we wondered about the role of CLIP-1[CLIP170]. Surprisingly, neither treatment with *clip-1(RNAi)* nor crossing with a strain carrying the *clip-1(gk470)* null mutation resulted in any significant alteration of the track densities (Fig. 4b). Similarly, crossing the null mutant with a centrosome-labelled strain carrying γTUB::GFP also resulted in no significant reduction of anaphase oscillation amplitudes (Fig. S4b and Supplementary Text 2.4). Anaphase oscillations were instrumental in this, as they are very sensitive to even a mild decrease in the number of active force generators (Supplementary Text 2.4). We concluded that dynein probably accumulates at the microtubule plus-ends via EBP-2[EB1], a hub for proteins localised at the microtubule plus-ends[75]. This is done with the help of dynactin and LIS-1, which resembles the findings in mammal protein experiments[76]. CLIP-1[CLIP170] promotes the hierarchical interaction *in vitro,* but surprisingly it may not have a role in nematodes.

We next investigated the link between dynein plus-end accumulation and cortical pulling force generation. We found that EBP-2[EB1] (but not its paralogs EBP-1 and EBP-3) moderately contributes to force generation (Supplementary Text 2.4 and Fig. S4c). We wondered whether this could be caused by dynein targeting at the cortex. We first partially depleted dynein in *ebp-2(gk756)* null mutant with labelled centrosomes by treating with *dyci-1(RNAi),* comparing the oscillation amplitude with control. We found no significant difference between that and the control treated by *dyci-1(RNAi)* under the same conditions (Fig. S4c), suggesting that EBP-2[EB1] and DYCI-1 are likely to act on the same pathway. We then analysed the cortical DYCI-1::mCherry spots on depleting EBP-2 either by *ebp-2(RNAi)* or by crossing to null *ebp-2* mutant, finding a drastic reduction in cortical track densities (Fig. S5f-i, k), similar to the one in the LSP. Overall, this suggests that dynein accumulated at microtubule plus-ends might be offloaded to the cortex, as is the case in budding yeast[8,23]. We did not, however, focus on the molecular details of this process. To summarize, dynein accumulation at microtubule plus-ends probably contributes, although moderately, to generating the cortical pulling forces.

Because RNAi depletion of EBP-2 only modestly decreased cortical forces, we next investigated the EBP-2–independent forces. We cannot totally exclude the idea that they are generated independently of dynein[80]. But since a mild depletion of the dynein subunit cancels out oscillations[21] (Fig. S4a) and broadly cortical pulling forces as suggested previously by the invagination assay[42], it is unlikely that a force generated independently of dynein would be sufficient to account for the close to normal forces observed upon *ebp-2(RNAi)*. We therefore looked for a second mechanism for dynein targeting at the cortex. In budding yeast, Bik1p/CLIP170 can target dynein independently of Bim1p/EB1[31]. However, oscillations measured in *ebp-2(0) clip-1$^{CLIP170}$(RNAi)* phenocopied the *ebp-2(0)* ones (Fig. S4c, *p=0.3*), with no additive phenotype. We therefore suggest that CLIP-1 plays no role in the secondary mechanism. Since the kinesin Kip2p (which has no known homolog in *C. elegans*) transports dynein in budding yeast, we investigated kinesin involvement in nematodes. However, none of the kinesins, whose depletion by RNAi decreases oscillation amplitudes (Fig. S4a), prevent dynein accumulation at the microtubule plus-ends (Fig. S5l and Supplementary Text 2.4). In conclusion, a functional approach suggested that EBP-2$^{EB1}$ moderately contributes to the targeting of dynein to the cortex, where it generates pulling forces. A second mechanism, independent of kinesins, account for the majority of these forces, likely bringing dynein to the cortex although we cannot entirely rule out the dynein-independent generation of some forces.

### Dynamics of DYCI-1::mCherry at the cell cortex: a larger dynein–microtubule attachment rate on the posterior side probably accounts for the imbalance in cortical pulling forces.

The positioning of the spindle during asymmetric division relies on an imbalance in cortical forces[38], with about double the *active* force generators on the posterior than on the anterior cortices[40]. The simplest way for this imbalance to occur is to have asymmetrical numbers of total force generators, sum of those that are inactive/unbound and those active, bound to a microtubule and pulling[21] (Fig. 6 top). On the whole, there are four possible causes of force imbalances. Possibility 1 is that there is a higher total dynein count on the posterior side, as explained above. In possibility 2, on the posterior side there are more dyneins that assemble in the trimeric complex with GPR-1/2 and LIN-5[41], and more microtubules attaching to this complex to engage in pulling (Fig. 6 middle). In accounting for the anaphase oscillations, we collectively modelled both the dynein binding rate to the cortex in the trimeric complex, and its subsequent attachment to microtubules, with an effective binding-rate termed on-rate[21]. In possibility 3, dynein may spend more time pulling on the posterior side microtubules (Fig. 6 bottom). This persistence pulling, termed processivity, is the inverse of the detachment– or off–rate, and has been suggested to cause the increase in forces during anaphase[21]. Tracking dynein dynamics at the cortex should nicely test these three hypotheses. Finally, possibility 4 is that there is a differential regulation of the intrinsic properties of dynein when generating forces, for instance the detachment sensitivity to force, the maximum velocity or the stall force. We refuted this fourth possibility because it does not result in an asymmetric number of active force generators[40] (Fig. 5c,d green curves, see below), and because the comparison of the anterior and posterior centrosomal oscillations during anaphase is hardly compatible with this model (Supplementary Text 2.5). We thus used our assay to explore the first three possibilities, which are related to dynein total counts and dynamics.

To test whether the force imbalance results from asymmetry in total numbers, we viewed DYCI-1::mCherry at the cortex (Fig. 1c and Movie S3). We analysed the dynamics at the cortex and, found equal proportions of directed and diffusive-like tracks (Fig. 2b,e).

We consistently found that diffusive-like (anisotropic) tracks displayed a diffusive motion in the BCA model classification (Fig. 2g). Furthermore, both spot types resided at the cortex during less than 1 s (Fig. 2c,d and Supplementary Text 2.6), consistent with the "tug-of-war" model's predictions[21]. Since dynein spots mostly display a directed motion in the cytoplasm, we reasoned that directed tracks might correspond to dynein spots that are completing their arrival on a microtubule plus-end at the cortex because the optical sectioning made by spinning disk microscopy allowed viewing of the sub-cortical region. Consistent with this idea, BCA analysis applied to the directed tracks resulted in motion probabilities that were similar to those obtained from the LSP (Fig. 2f). To test whether directed tracks are related to microtubule growth sub-cortically towards the cortex, we used RNAi to deplete EFA-6, a putative microtubule regulator[81]. As expected, this resulted in directed tracks that were longer at the cortex (Fig. 5a, middle). Directed tracks were more numerous than in the control, while the diffusive-like population was not significantly affected (Fig. 5a,b). We therefore suggest that directed tracks are likely to mainly correspond to dynein at the plus ends of microtubules that are arriving at the cortex. We showed above that at the cortex, dynein spots colocalized partially with GPR-1::YFP. We thus hypothesized that the other type of dynein spots at the cortex, diffusive-like, may correspond to dynein that generates pulling forces. Because GPR-1/2 is involved in force generation[21,39,40] and is furthermore posited to be its limiting factor[46,47], we reasoned that its depletion should not alter the directed counts, mostly affecting the diffusive ones. And in fact when we measured the number of dynein tracks in embryo subjected to a partial *gpr-1/2(RNAi)*, we found important differences only in the diffusive population (Fig. 5a, middle and 5c,d), confirming that this is the one involved in force generation. Importantly, differential interference contrast (DIC) microscopy measurements of the posterior centrosome oscillations showed that they disappeared in 7 out of 7 embryos, while present in 8 out of 8 control embryos, confirming the penetrance of the RNAi. This hypomorphic treatment preserved a posterior displacement indistinguishable from control with a final posterior centrosome position at 76 ± 7% of embryo length (mean ± SD) upon *gpr-1/2(RNAi)* compared to 79 ± 1% in control, and thus a normal positional regulation of forces[47,82].

We then focused on the diffusive population to uncover the mechanism creating the cortical force imbalance. We began by considering possibility 1 that of an asymmetry in total force generator counts (Fig. 6 top). We observed dynein tracks within four equal regions (Fig. 5e), and measured the track densities (green lines, Fig. 5c,d). Since force generation is said to be reduced in the middle region[83] (which corresponds to the LET-99 domain[84]), we focused on the regions at the tips, indicated by numbers 1 and 4. In non-treated embryos, we compared the posterior and anterior tip track counts (regions 1 and 4, respectively) using a two states model comparing probability for a microtubule to contact in region 4 versus 1, and found a very significant higher count on posterior tip compared to anterior one (Fig. 5c,d and Supplementary Text 2.8). This is compatible with both possibilities 1 and 2. If the total number at the cortex is asymmetric (possibility 1), one would expect a stronger targeting of dynein to the cortex in the posterior half. Because dynein accumulation at the microtubule plus-ends may contribute to the cortical pool, we computed the posterior-to-anterior ratio of the tracks in the LSP. This yielded 0.95 ± 0.09 for the directed tracks, and 1.0 ± 0.1 for the diffusive-like ones ($N$ = 7 embryos, 1341 tracks), which hardly seems compatible with asymmetrical dynein targeting. Since there is a second, probably microtubule-independent, mechanism that contributes to the targeting, we used FCS to measure the concentration of dynein in the embryo halves, and the results were not asymmetrical (Fig. S5m and Supplementary Text 1.4). This suggests that an equal flux of dynein is likely to reach both parts of the cortex, so the force imbalance is probably not caused by an asymmetry in total numbers. To be sure, we performed a functional assay using anaphase oscillations. Asymmetrical dynein counts would result in a larger amplitude but a smaller frequency on the posterior side compared to anterior one (Supplementary Text 2.5). This is not consistent with the measurements, which showed both larger frequency and amplitude on posterior side. Altogether, we suggest that possibility 1, which was that the total number of dyneins (active and inactive) would be asymmetric, is unlikely (Fig. 6, top).

We next considered the dynamics of force generators: possibilities 2 and 3 which cover the attachment and detachment rate, respectively (Fig. 6 middle, bottom). The detachment or off-rate is the inverse of the processivity, and is said to reflect mitotic progression[21,85]. The asymmetry in density (Fig. 5c,d) is rather compatible with possibility 2, however, to further test whether processivity also reflects polarity, we measured the residency time (duration of the tracks), within the four previously defined regions. We found no difference between the posterior and anterior halves (Fig. 5f,g). We concluded that the off-rate does not account for the force imbalance. We therefore suggested that dynein binds with a higher rate or affinity on the posterior side than the anterior one, corresponding to possibility 2 which proposes an asymmetry in the on-rate. To ascertain this suggestion, using RNAi, we partially suppressed GPR-1/2, known to decrease force imbalance[21,39,40] and found a lower track density, especially on the posterior tip (Fig. 5c). We also noticed that an asymmetric on-rate correctly accounts for the different characteristics of the anterior and posterior centrosomal oscillations (Supplementary Text 2.5). We concluded that in the *C. elegans* zygote, dynein attachment rate is likely larger on the posterior cortex, accounting for the force imbalance that reflects polarity in this asymmetric division, displacing the spindle posteriorly during anaphase[21,40].

# Discussion

Using a fluorescence-labelled dynein DYCI-1 subunit as a *bona fide* dynein reporter, we developed a method for the analysis of dynein dynamics and applied it in both the cytoplasm and the cell cortex. In the cytoplasmic LSP, we found that dynein accumulates at the microtubule plus-ends in a manner dependent on EBP-2$^{EB1}$, dynactin, and LIS-1[50]. This mechanism shows striking similarities with findings based on *in vitro* assays with human proteins[76], which suggests a conservation across evolution. Commonalities with budding yeast[8,22-29] are less numerous as, in contrast to this one, dynactin or EBP-2$^{EB}$ was required in nematode. In particular, a hierarchical interaction has been proposed in human wherein dynactin permits dynein to indirectly hitchhike on EB1[86], and our results are fully compatible with this model. This contrasts with CLIP-1$^{CLIP170}$, whose homolog facilitates hitchhiking in a human protein-based minimal system[86]. In nematodes, this protein appears not to have a role in dynein accumulation at the microtubule plus-ends. This might be explained by the weak homology between CLIP-1 and mammalian CLIP170[87]. CLIP-1 could instead be a tubulin-folding cofactor B[88], and to date no other homolog has been predicted. In budding yeast, Bik1p/CLIP170 is indispensable: together with the Kip2p kinesin it contributes to dynein targeting[25,27,28], and is also able to track plus-ends on its own and compensate for the lack of Bim1p/EB1[25,29,31].

At the plus-ends dynein displays attachment/detachment dynamics similar to EBP-2. This means that dynein is not actively transported towards the cell periphery, and is instead briefly immobilized on the microtubule lattice. It is therefore puzzling that such a mechanism could contribute to targeting dynein to the cortex, in turn generating cortical forces, even if only modestly. There are two advantages of concentrating dynein at microtubule plus-ends. First, in order to generate pulling forces there needs to be a meeting between multiple players, including a microtubule, a dynein-dynactin complex, GPR-1/2, and LIN-5. These have to create the force-generating complex[41] during dynein's brief residency time. In this respect, it is useful to have dynein-dynactin and microtubules already brought together, and further studies are needed to elucidate the details of offloading. The second advantage is that it may contribute to bringing dynein to the cortex by biasing its diffusion towards the periphery. Indeed, when dynein detaches at the back of the microtubule plus-end GTP cap along with EBP-2 and other accessory proteins, its affinity for the EBP-2 bound to this cap may favour its diffusion towards the microtubule plus-ends[89], i.e. the cell periphery. The sort of biased diffusion exemplified here[90] has been discussed in the context of molecular motors, particularly ones having one motor domain and being processive, although the mechanisms would require further investigations.

Why were cortical forces partially preserved upon EBP-2$^{EB1}$ suppression even with the disappearance of plus-end accumulation? The possibility that cortical forces position the spindle independently of dynein has been investigated using hypomorphic or temperature-sensitive alleles[80], but some dynein activity might have remained. Oscillation disappears and posterior displacement is strongly reduced upon partial *dli-1(RNAi)*[21,41], suggesting that dynein contributes for most if not all cortical pulling force generation. Therefore we hypothesized that another mechanism also targets dynein to the cortex. Because FCS measurements show dynein attachment and detachment dynamics at the microtubule plus-ends that are the same as EBP-2, it is unlikely that dynein is transported along the microtubule lattice by kinesins or via one-dimensional diffusion along its lattice, as the resulting dynamics would follow the "antenna model," i.e. a microtubule length dependent plus-end accumulation[91,92]. The most plausible alternative mechanism is therefore a three-dimensional dynein diffusion, making sense because of the large cytoplasmic concentrations of 177 ± 60 nM (32 ± 11 molecules in 0.3 fl FCS focal volume, $N$ = 8 embryos, 38 spots; Supplementary Text 1.4). We can estimate that about 30 dynein molecules per second arrive at a cortex half (Supplementary Text 2.7)[93]. This result is a bit low, but still relevant[40]. Interestingly, it has been suggested that the LIN-5 homolog of NuMA, part of the cortical force-generating complex, recruits dynein at the cortex independently of astral microtubules[59].

How do these dynein targeting mechanisms contribute to cortical force generation? We found two populations of dynein at the cortex. One has a directed motion, residing longer at the cortex, and represents microtubule plus-ends that are finishing their approach to the cortex. The second stays for less time and displays diffusive-like motion, which may correspond to pulling force-generating events. In addition to having densities which reflect the expected force imbalances and GPR-1/2 dependencies, our observations are supported by two additional indications. First, the observed residency time is consistent with the brief cortical stay measured for microtubules[82,94] and with the estimated force generator run-times produced in anaphase centrosomal oscillation models[21]. Second, the number of spots is consistent with the expected active force generator count of 10 to 100 per cortex half[40,42]. Indeed, in the posterior section, we observed about 0.008 diffusive tracks per μm$^2$ of visible cortex area (instantaneous density), which is about 20 diffusive tracks in the posterior half at any instant. We also assessed as unlikely the possibility that diffusive tracks corresponded to lateral diffusion of microtubule plus-end "searching" for force generator anchoring[6], because diffusive-like tracks are independent of microtubule growth, as probed by *efa-6(RNAi)*[81], but are dependent on the member of cortical force–generating complex GPR-1/2[41,46,47]. When analysing track densities across four equal regions, we obtained the highest densities in the two central ones, which is consistent with the higher density of microtubule contacts at the cortex closest to the centrosomes[95]. However there was no asymmetry in the densities or residency times between the central anterior and posterior regions, likely due to LET-99 inhibition of force generation in a band extending from 40 to 70% of the anteroposterior axis[83]. Therefore, these two middle regions cannot contribute to force imbalance. Although some tracks that have a mix of directed (microtubules arriving at the cortex) and diffusive-like motions were classified according to which class they spent the longest time in, investigation of the diffusive population still offers a unique opportunity for deciphering the details of polarity-induced force imbalances. As with many other asymmetric divisions[19,20,36,37], asymmetric spindle positioning during the division of the *C. elegans* zygote relies on a cortical force imbalance, with stronger forces arrayed on the posterior side[21,40]. In the nematode, this is mediated by the GPR-1 and -2 proteins, along with dynein[41], both part of the force generating complex which is more concentrated at the posterior tip of the cortex. GPR-1/2 was reported to be enriched on posterior side[46,47], therefore dynein's increased attachment or on-rate at the posterior cortex can be seen as displacing the association/dissociation reaction to trimeric complex assembly by increasing the concentration of one reactant. Meanwhile, an identical number of dynein arrive at the cortex from the microtubule plus-ends or the cytoplasm (Fig. 6). As a result, the density of dynein tracks involved in force generation (diffusive-like) is increased. In contrast, such a mechanism will not lead to an increase in dynein residency time at the cortex, which is related to the load-dependent detachment rate of the dynein[21] and to an external regulation probably related to mitotic progression[85]. We suspect that the GPR-1/2 dependent increase in dynein residency time (non polarized) is indirect, since GPR-1/2 would be needed to regulate the localization or activation of a member of the force-generating complex which in turn regulates processivity. Overall, the proposed mechanism to build force imbalance is perfectly in line with the fast dynein turnover at the cortex. With a residency time below 1 s, localization can be adapted according to internal evolution and dynamic polarity cues[47,96,97].

Dynein's high levels of dynamics at the cell cortex require a very efficient targeting mechanism. Cytoplasmic diffusion is reinforced by the accumulation of dynein at the microtubule plus-ends, where it indirectly hitches a ride with EBP-2 with the help of dynactin and LIS-1. These dynein dynamics at the cell cortex revealed that the polarity-based asymmetry in number of active force generators is likely due to an increased dynein binding rate on the posterior side, probably enabled by higher amounts of GPR-1/2[46,47]. This control is part of the threefold regulation of the forces that position the mitotic spindle: through polarity as reported here, through positioning of the posterior spindle pole[47,95], and through the force generator persistence to pull on microtubules (processivity)[21]. We report here that this processivity, the inverse of the off-rate, is not polarized. It may rather reflect mitotic progression[85]. One-cell embryo division in the nematode has paved the way to understanding asymmetrical division mechanisms[36,37], and it would be very interesting to investigate whether polarized force is due to asymmetric force generator binding rate in other organisms.

# Material and Methods

**Culturing *C. elegans***

*C. elegans* strains were cultured as described in[98] and dissected to obtain embryos. All strains containing DYCI-1::mCherry were maintained at 25ºC, while functional experiments (anaphase oscillations) investigating the role of CLIP170, EB1 homologs, and kinesins were performed at 18ºC. The exception to this was clip-1(gk470), maintained at 23ºC, and the corresponding strains were cultured at the same temperature.

***C. elegans* strains used**

The Bristol strain N2 was used as the standard wild-type strain[98]. The following fluorescent strains were used: TH163 (DYCI-1::mCherry)[56]; TH27 (GFP::TBG-1)[99]; TH65 (YFP::TBA-2)[94]; TH66 (GFP::EBP-2)[79]; DE74 (GFP::PLCδ1-PH)[100]; TH110 (mCherry::PAR-6)[101]; and TH242 (GPR-1::YFP)[102]. Standard genetic crosses were done to generate these multi-labelled combinations: JEP2 (DYCI-1::mCherry,YFP::TBA-2); JEP12 (DYCI-1::mCherry,GFP::EBP-2); JEP20 (DYCI-1::mCherry,GFP::PLCδ1-PH) and JEP58 (DYCI-1::mCherry,YFP::GPR-1). To obtain JEP27 and JEP32 carrying the GFP::TBG-1 transgene and the *ebp-2(gk756)* or *clip-1(gk470)* mutations, we crossed TH27 with VC1614 or VC1071, respectively[103]. The strain carrying the *dyci-1(tm4732)* lethal mutation was provided by the Mitani Lab via the National BioResource Project and JEP9 was generated by crossing with VC2542 to balance the lethal mutation with nT1[qIS51] translocation. JEP30 and JEP40 strains homozygous for *dyci-1(tm4732)* were obtained by double-crossing JEP9 with JEP23 and TH163, respectively (Supplementary Text 1.2). The transgenes encoding the GFP, YFP, and mCherry fusion proteins in all constructs but DYCI-1::mCherry were under the control of the pie-1 promoter.

**Gene silencing by RNA interference**

Except when otherwise stated, embryonic RNAi was done by feeding, using both the Ahringer library[18,104] and clones ordered from Source BioScience. However, the clones for ebp-1/3 and klp-18 were made in the lab. To do this, N2 genomic DNA was used to amplify a region from the target gene (see Table 1). This was then cloned into the L4440 RNAi feeding vector and transformed into HT115 bacteria. For ebp-1, a region corresponding to exons 2 and 3 after splicing was amplified using four long primers and fused by PCR amplification before the L4440 cloning. The primers used for amplification are listed in Table 1.

| Target | Forward primer | Reverse primer |
|---|---|---|
| ebp-1/3 | 5' ACCGGGAGTCGATATGGC 3' | 5' TCAACATTTCCAATCGATTCATT 3' |
| ebp-1 | 5' TCGTCTTGAATTGGATTGGCTTTCCAACTGGAAACTAGTGCAGACTACGTGGAAGAATTT 3' | 5' TTGTCCTGAAATTTTCCCTTAATCAATTTATCAACAGGAATCACTTTCTCGACACCCAAATTCTTCCACGTAGTCTGCAC 3' |
|  | 5' GATTAAGGGAAAATTTCAGGACAACTTTGAATTCTTGCAA | 5' CATTACGTGCTTGCATTGGATCATACTCATGTCCATCATAGTTAGC |

| | TGGTTCAAGAAATTGTTCGATGCTAACTATGATGGACATGAGTATGA 3' | 3' |
| --- | --- | --- |
| klp-18 | 5' ACGGAATTCGCATCACAGTT 3' | 5' CAATCTGTTCGTTTTCTGATCC 3' |

**Table 1: Primers used in this study.**

For *ebp-1* and *ebp-1/3* RNAi treatment, we observed a 40-60% reduction in the number of transcripts as measured by Q-RT-PCR without changes in the *ebp-2* mRNA levels. Total RNA was extracted from around 20 worms using a Direct-Zol RNA MicroPrep kit (Zymo Research). Production of cDNA was done with a ProtoScript II First Strand cDNA Synthesis kit (New England Biolabs). For Q-PCR, Power SYBR Green PCR Master Mix (Thermo Fisher Scientific) was used with a 7900HT Fast Real-Time PCR System (Applied Biosystems).

To allow for the varied expression of DYCI-1::mCherry in the randomly integrated strain, each RNAi experiment was compared or normalized to non-treated embryos imaged on the same day (e.g. Fig. 4a,b, 5, and 6b-d).

Except where otherwise stated, RNA interference was partial: observation was performed 23-25h after plating the worms. In particular and to avoid too strong or unrelated phenotypes, we used the following treatment durations when observing the randomly integrated DYCI-1::mCherry strain (TH163): *lin-5(RNAi)*, 17h; *gpr-1/2(RNAi)*, 48h; *lis-1(RNAi)*, 16h-18h;, *klp-3(RNAi)*, 18h; *klp-7(RNAi)*, 18h; *dnc-1(RNAi)*, 16h-18h; and *ebp-2(RNAi)*, 20h. When using γTUB::GFP (TH27) to investigate oscillation, embryos treated by kinesins RNAi where observed after 24h; by the dynein subunit *dyci-1(RNAi)* after 16h; and by *dli-1(RNAi)* after 24h.

**Live imaging**

Embryos were dissected in M9 buffer and mounted on pads (2% w/v agarose, 0.6% w/v NaCl, 4% w/v sucrose). We imaged one-cell *C. elegans* embryos during metaphase and anaphase. Dynein/EBP-2 tracking was performed on a LEICA DMI6000/Yokogawa CSU-X1 M1 spinning disc microscope, using an HCX Plan APO 100x/ 1.40 Oil objective. A Fianium white light laser conveniently filtered around 488 nm and 561 nm by a homemade setup was used for illumination (patent pending[105]). Images were acquired with a 200 ms exposure time (5 Hz) using a Photometrics Evolve Camera (Roper) and MetaMorph software (Molecular Devices) without binning. During the experiments, the embryos were kept at 24°C. To image embryos at the LSP, we typically moved the focus between 3 and 5 μm below the spindle plane (Fig. S2d)[106].

**Image processing**

The standard deviation maps (SDM) were generated with Fiji's ZProject plugin for ImageJ, specifying a "standard deviation" over 6 s of the time-lapse image sequence[57,107].

The tracking of labelled centrosomes and analysis of trajectories were performed by a custom tracking software[21] and developed using Matlab (The MathWorks). Tracking of -20°C methanol-fixed γTUB::GFP embryos indicated an accuracy to 10 nm. Embryo orientations and centres were obtained by cross-correlation of embryo background cytoplasmic fluorescence with artificial binary images mimicking embryos, or by contour detection of the cytoplasmic membrane using background fluorescence of cytoplasmic

γTUB::GFP with the help of an active contour algorithm[108]. The results were averaged over all of the replicas for each condition.

**Statistics**

The displayed centre values are the means except when otherwise stated. Averaged values were compared using a two-tailed Student's *t*-test with the Welch-Satterthwaite correction for unequal variance, except if stated otherwise. For the sake of simplicity, we encoded confidence level using stars as follows: ♦, $p < 0.1$; *, $p \leq 0.05$; **, $p \leq 0.01$; ***, $p \leq 0.001$; ****, $p \leq 0.0001$; and n.s., non-significant, $p > 0.1$. The n.s. indication may be omitted for clarity's sake. We abbreviated standard deviation (S.D.); standard error (s.e.); and standard error of the mean (s.e.m.).

**Code and data availability**

The computer codes and datasets generated and analysed during this study are available upon reasonable request from the corresponding author.

# Acknowledgements

We thank Prof. Anthony A. Hyman and Dr Mihail Sarov for providing strains, in particular the kind gift of the randomly integrated DYCI-1::mCherry TH163; Dr J.W. Dennis for the DE74 strain; Dr S. Redemann for preliminary data on this project; Drs N. Monnier and M. Bathes for support with Bayesian analysis; Dr G. Michaux for the feeding clones library; and Drs B. Mercat, A. Pacquelet, X. Pinson, H. Bouvrais, Y. Le Cunff, G. Michaux, R. Le Borgne, and S. Huet for technical help, critical comments on the manuscript, and discussions about the project. RRG and JP were supported by a CNRS ATIP starting grant and La Ligue Nationale Contre le Cancer. Some strains were provided by the CGC, which is funded by the NIH Office of Research Infrastructure Programs (P40 OD010440; University of Minnesota, USA), by the National BioResource Project (Tokyo University, Japan), and by the *C. elegans* Gene Knockout Consortium. The MosSci strain was made by the UMS 3421 Biology of *Caenorhabditis elegans* facility, CNRS/UCBL (Lyon, France). Microscopy imaging was performed at the MRIC facility, UMS 3480 CNRS/US 18 INSERM/University of Rennes 1. The FCS microscopy setup was funded by ARC grant #EML20110602452, and the spinning disk microscopy was funded by the CNRS, Rennes Métropole, and Region Bretagne (grant AniDyn-MT). We also acknowledge *plan cancer* grant BIO2013-02 and COST EU action BM1408 (GENiE).

# Author contributions

RRG, LC, and JP designed the research, analysed the data and wrote the paper. RRG, LC, and SP performed the research. JR and MT contributed new analytic tools.

# References

1. Gonczy P, Pichler S, Kirkham M, Hyman AA. Cytoplasmic dynein is required for distinct aspects of MTOC positioning, including centrosome separation, in the one cell stage Caenorhabditis elegans embryo. *J Cell Biol* **147**, 135-150 (1999).
2. Dujardin DL, Vallee RB. Dynein at the cortex. *Current opinion in cell biology* **14**, 44-49 (2002).
3. Nguyen-Ngoc T, Afshar, K., and Gonczy, P. Coupling of cortical dynein and G alpha proteins mediates spindle positioning in Caenorhabditis elegans. Nat. Cell Biol. *9* **1294-1302**, (2007).
4. Karki S, Holzbaur EL. Cytoplasmic dynein and dynactin in cell division and intracellular transport. *Current opinion in cell biology* **11**, 45-53 (1999).
5. Kotak S, Gonczy P. Mechanisms of spindle positioning: cortical force generators in the limelight. *Current opinion in cell biology* **25**, 741-748 (2013).
6. Laan L*, et al.* Cortical dynein controls microtubule dynamics to generate pulling forces that position microtubule asters. *Cell* **148**, 502-514 (2012).
7. Carminati JL, Stearns T. Microtubules orient the mitotic spindle in yeast through dynein-dependent interactions with the cell cortex. *J Cell Biol* **138**, 629-641 (1997).
8. Markus SM, Lee WL. Regulated offloading of cytoplasmic dynein from microtubule plus ends to the cortex. *Dev Cell* **20**, 639-651 (2011).
9. Moore JK, Li J, Cooper JA. Dynactin function in mitotic spindle positioning. *Traffic* **9**, 510-527 (2008).
10. Shaw SL, Yeh E, Maddox P, Salmon ED, Bloom K. Astral microtubule dynamics in yeast: a microtubule-based searching mechanism for spindle orientation and nuclear migration into the bud. *J Cell Biol* **139**, 985-994 (1997).
11. Collins ES, Balchand SK, Faraci JL, Wadsworth P, Lee WL. Cell cycle-regulated cortical dynein/dynactin promotes symmetric cell division by differential pole motion in anaphase. *Mol Biol Cell* **23**, 3380-3390 (2012).
12. McNally FJ. Mechanisms of spindle positioning. *J Cell Biol* **200**, 131-140 (2013).
13. King SM. *Dyneins : structure, biology and disease*, 1st edn. Academic Press (2012).
14. Pfister KK, Lo KW-H. Cytoplasmic Dynein Function Defined by Subunit Composition. In: *Dyneins : structure, biology and disease* (ed King SM). 1st edn. Academic Press (2012).
15. Pfister KK*, et al.* Genetic analysis of the cytoplasmic dynein subunit families. *PLoS Genet* **2**, e1 (2006).
16. King SM. AAA domains and organization of the dynein motor unit. *J Cell Sci* **113**, 2521-2526 (2000).
17. Sonnichsen B*, et al.* Full-genome RNAi profiling of early embryogenesis in Caenorhabditis elegans. *Nature* **434**, 462-469 (2005).
18. Kamath RS, Ahringer J. Genome-wide RNAi screening in Caenorhabditis elegans. *Methods* **30**, 313-321 (2003).
19. McNally FJ. Mechanisms of spindle positioning. *J Cell Biol* **200**, 131-140 (2013).
20. di Pietro F, Echard A, Morin X. Regulation of mitotic spindle orientation: an integrated view. *EMBO Rep* **17**, 1106-1130 (2016).
21. Pecreaux J*, et al.* Spindle oscillations during asymmetric cell division require a threshold number of active cortical force generators. *Curr Biol* **16**, 2111-2122 (2006).
22. Moore JK, Stuchell-Brereton MD, Cooper JA. Function of dynein in budding yeast: mitotic spindle positioning in a polarized cell. *Cell Motil Cytoskeleton* **66**, 546-555 (2009).

23. Lee WL, Kaiser MA, Cooper JA. The offloading model for dynein function: differential function of motor subunits. *J Cell Biol* **168**, 201-207 (2005).
24. Sheeman B, *et al.* Determinants of S. cerevisiae dynein localization and activation: implications for the mechanism of spindle positioning. *Curr Biol* **13**, 364-372 (2003).
25. Carvalho P, Gupta ML, Jr., Hoyt MA, Pellman D. Cell cycle control of kinesin-mediated transport of Bik1 (CLIP-170) regulates microtubule stability and dynein activation. *Dev Cell* **6**, 815-829 (2004).
26. Lee WL, Oberle JR, Cooper JA. The role of the lissencephaly protein Pac1 during nuclear migration in budding yeast. *J Cell Biol* **160**, 355-364 (2003).
27. Roberts AJ, Goodman BS, Reck-Peterson SL. Reconstitution of dynein transport to the microtubule plus end by kinesin. *eLife* **3**, e02641 (2014).
28. Markus SM, Punch JJ, Lee WL. Motor- and tail-dependent targeting of dynein to microtubule plus ends and the cell cortex. *Curr Biol* **19**, 196-205 (2009).
29. Markus SM, Plevock KM, St Germain BJ, Punch JJ, Meaden CW, Lee WL. Quantitative analysis of Pac1/LIS1-mediated dynein targeting: Implications for regulation of dynein activity in budding yeast. *Cytoskeleton (Hoboken)* **68**, 157-174 (2011).
30. Ananthanarayanan V, Schattat M, Vogel SK, Krull A, Pavin N, Tolic-Norrelykke IM. Dynein motion switches from diffusive to directed upon cortical anchoring. *Cell* **153**, 1526-1536 (2013).
31. Caudron F, Andrieux A, Job D, Boscheron C. A new role for kinesin-directed transport of Bik1p (CLIP-170) in Saccharomyces cerevisiae. *J Cell Sci* **121**, 1506-1513 (2008).
32. Kobayashi T, Murayama T. Cell cycle-dependent microtubule-based dynamic transport of cytoplasmic dynein in mammalian cells. *PLoS ONE* **4**, e7827 (2009).
33. Splinter D, *et al.* BICD2, dynactin, and LIS1 cooperate in regulating dynein recruitment to cellular structures. *Mol Biol Cell* **23**, 4226-4241 (2012).
34. Molina-Calavita M, Barnat M, Elias S, Aparicio E, Piel M, Humbert S. Mutant huntingtin affects cortical progenitor cell division and development of the mouse neocortex. *J Neurosci* **34**, 10034-10040 (2014).
35. Bieling P, *et al.* Reconstitution of a microtubule plus-end tracking system in vitro. *Nature* **450**, 1100-1105 (2007).
36. Morin X, Bellaiche Y. Mitotic spindle orientation in asymmetric and symmetric cell divisions during animal development. *Dev Cell* **21**, 102-119 (2011).
37. Gonczy P. Mechanisms of asymmetric cell division: flies and worms pave the way. *Nat Rev Mol Cell Biol* **9**, 355-366 (2008).
38. Grill SW, Gonczy P, Stelzer EH, Hyman AA. Polarity controls forces governing asymmetric spindle positioning in the Caenorhabditis elegans embryo. *Nature* **409**, 630-633 (2001).
39. Colombo K, Grill SW, Kimple RJ, Willard FS, Siderovski DP, Gonczy P. Translation of polarity cues into asymmetric spindle positioning in Caenorhabditis elegans embryos. *Science* **300**, 1957-1961 (2003).
40. Grill SW, Howard J, Schaffer E, Stelzer EH, Hyman AA. The distribution of active force generators controls mitotic spindle position. *Science* **301**, 518-521 (2003).
41. Nguyen-Ngoc T, Afshar K, Gonczy P. Coupling of cortical dynein and G alpha proteins mediates spindle positioning in Caenorhabditis elegans. *Nat Cell Biol* **9**, 1294-1302 (2007).
42. Redemann S, *et al.* Membrane invaginations reveal cortical sites that pull on mitotic spindles in one-cell C. elegans embryos. *PLoS One* **5**, e12301 (2010).
43. Rose L, Gonczy P. Polarity establishment, asymmetric division and segregation of fate determinants in early C. elegans embryos. *WormBook*, 1-43 (2014).
44. Srinivasan DG, Fisk RM, Xu H, van den Heuvel S. A complex of LIN-5 and GPR proteins regulates G protein signaling and spindle function in C elegans. *Genes Dev* **17**, 1225-1239 (2003).

45. Galli M, *et al.* aPKC phosphorylates NuMA-related LIN-5 to position the mitotic spindle during asymmetric division. *Nat Cell Biol* **13**, 1132-1138 (2011).
46. Park DH, Rose LS. Dynamic localization of LIN-5 and GPR-1/2 to cortical force generation domains during spindle positioning. *Dev Biol* **315**, 42-54 (2008).
47. Riche S, Zouak M, Argoul F, Arneodo A, Pecreaux J, Delattre M. Evolutionary comparisons reveal a positional switch for spindle pole oscillations in Caenorhabditis embryos. *J Cell Biol* **201**, 653-662 (2013).
48. Cockell MM, Baumer K, Gonczy P. lis-1 is required for dynein-dependent cell division processes in C. elegans embryos. *J Cell Sci* **117**, 4571-4582 (2004).
49. Splinter D, *et al.* BICD2, dynactin and LIS1 cooperate in regulating dynein recruitment to cellular structures. *Mol Biol Cell*, (2012).
50. Jha R, Roostalu J, Trokter M, Surrey T. Combinatorial Regulation Of The Balance Between Dynein Microtubule End Accumulation And Initiation Of Directed Motility. *bioRxiv*, 126508 (2017).
51. Trokter M, Surrey T. LIS1 Clamps Dynein to the Microtubule. *Cell* **150**, 877-879 (2012).
52. Lammers LG, Markus SM. The dynein cortical anchor Num1 activates dynein motility by relieving Pac1/LIS1-mediated inhibition. *J Cell Biol* **211**, 309-322 (2015).
53. Williams SE, Ratliff LA, Postiglione MP, Knoblich JA, Fuchs E. Par3-mInsc and Galphai3 cooperate to promote oriented epidermal cell divisions through LGN. *Nat Cell Biol* **16**, 758-769 (2014).
54. Moore JK, Cooper JA. Coordinating mitosis with cell polarity: Molecular motors at the cell cortex. *Semin Cell Dev Biol* **21**, 283-289 (2010).
55. Sarov M, *et al.* A genome-scale resource for in vivo tag-based protein function exploration in C. elegans. *Cell* **150**, 855-866 (2012).
56. Sarov M, *et al.* A recombineering pipeline for functional genomics applied to Caenorhabditis elegans. *Nat Methods* **3**, 839-844 (2006).
57. Rostampour AR, Reeves AP, Mitchell OR. Use of temporal variance for moving object extraction. In: *Computers and Communications, 1988. Conference Proceedings., Seventh Annual International Phoenix Conference on*) (1988).
58. Gonczy P, Pichler S, Kirkham M, Hyman AA. Cytoplasmic dynein is required for distinct aspects of MTOC positioning, including centrosome separation, in the one cell stage Caenorhabditis elegans embryo. *J Cell Biol* **147**, 135-150 (1999).
59. Schmidt R, Fielmich LE, Grigoriev I, Katrukha EA, Akhmanova A, van den Heuvel S. Two populations of cytoplasmic dynein contribute to spindle positioning in C. elegans embryos. *J Cell Biol*, (2017).
60. Boulin T, Bessereau JL. Mos1-mediated insertional mutagenesis in Caenorhabditis elegans. *Nat Protoc* **2**, 1276-1287 (2007).
61. Robert V, Bessereau JL. Targeted engineering of the Caenorhabditis elegans genome following Mos1-triggered chromosomal breaks. *Embo J* **26**, 170-183 (2007).
62. Dragestein KA, *et al.* Dynamic behavior of GFP-CLIP-170 reveals fast protein turnover on microtubule plus ends. *J Cell Biol* **180**, 729-737 (2008).
63. Grill SW, Hyman AA. Spindle positioning by cortical pulling forces. *Dev Cell* **8**, 461-465 (2005).
64. Jaqaman K, *et al.* Cytoskeletal control of CD36 diffusion promotes its receptor and signaling function. *Cell* **146**, 593-606 (2011).
65. Yoder JH, Han M. Cytoplasmic dynein light intermediate chain is required for discrete aspects of mitosis in Caenorhabditis elegans. *Mol Biol Cell* **12**, 2921-2933 (2001).
66. Boxem M, *et al.* A protein domain-based interactome network for C. elegans early embryogenesis. *Cell* **134**, 534-545 (2008).
67. Li S, *et al.* A map of the interactome network of the metazoan C. elegans. *Science* **303**, 540-543 (2004).
68. Trokter M, Mucke N, Surrey T. Reconstitution of the human cytoplasmic dynein complex. *Proc Natl Acad Sci U S A* **109**, 20895-20900 (2012).

69. Shivaraju M, Unruh JR, Slaughter BD, Mattingly M, Berman J, Gerton JL. Cell-cycle-coupled structural oscillation of centromeric nucleosomes in yeast. *Cell* **150**, 304-316 (2012).
70. Coupe P, Munz M, Manjon JV, Ruthazer ES, Collins DL. A CANDLE for a deeper in vivo insight. *Med Image Anal* **16**, 849-864 (2012).
71. Sage D, Neumann FR, Hediger F, Gasser SM, Unser M. Automatic tracking of individual fluorescence particles: application to the study of chromosome dynamics. *IEEE Trans Image Process* **14**, 1372-1383 (2005).
72. Jaqaman K, *et al.* Robust single-particle tracking in live-cell time-lapse sequences. *Nat Methods* **5**, 695-702 (2008).
73. Huet S, Karatekin E, Tran VS, Fanget I, Cribier S, Henry JP. Analysis of transient behavior in complex trajectories: application to secretory vesicle dynamics. *Biophysical journal* **91**, 3542-3559 (2006).
74. Monnier N, Guo SM, Mori M, He J, Lenart P, Bathe M. Bayesian approach to MSD-based analysis of particle motion in live cells. *Biophysical journal* **103**, 616-626 (2012).
75. Akhmanova A, Steinmetz MO. Tracking the ends: a dynamic protein network controls the fate of microtubule tips. *Nat Rev Mol Cell Biol* **9**, 309-322 (2008).
76. Duellberg C, Trokter M, Jha R, Sen I, Steinmetz MO, Surrey T. Reconstitution of a hierarchical +TIP interaction network controlling microtubule end tracking of dynein. *Nat Cell Biol*, (2014).
77. Schroer TA. Dynactin. *Annu Rev Cell Dev Biol* **20**, 759-779 (2004).
78. Skop AR, White JG. The dynactin complex is required for cleavage plane specification in early Caenorhabditis elegans embryos. *Curr Biol* **8**, 1110-1116 (1998).
79. Srayko M, Kaya A, Stamford J, Hyman AA. Identification and characterization of factors required for microtubule growth and nucleation in the early C. elegans embryo. *Dev Cell* **9**, 223-236 (2005).
80. Schmidt DJ, Rose DJ, Saxton WM, Strome S. Functional analysis of cytoplasmic dynein heavy chain in Caenorhabditis elegans with fast-acting temperature-sensitive mutations. *Mol Biol Cell* **16**, 1200-1212 (2005).
81. O'Rourke SM, Christensen SN, Bowerman B. Caenorhabditis elegans EFA-6 limits microtubule growth at the cell cortex. *Nat Cell Biol* **12**, 1235-1241 (2010).
82. Bouvrais H, Chesneau L, Pastezeur S, Delattre M, Pecreaux J. Astral microtubule dynamics regulate anaphase oscillation onset and set a robust final position of the C. elegans zygote spindle. *bioRxiv*, 103937 (2017).
83. Krueger LE, Wu JC, Tsou MF, Rose LS. LET-99 inhibits lateral posterior pulling forces during asymmetric spindle elongation in C. elegans embryos. *J Cell Biol* **189**, 481-495 (2010).
84. Wu JC, Rose LS. PAR-3 and PAR-1 inhibit LET-99 localization to generate a cortical band important for spindle positioning in Caenorhabditis elegans embryos. *Mol Biol Cell* **18**, 4470-4482 (2007).
85. McCarthy Campbell EK, Werts AD, Goldstein B. A cell cycle timer for asymmetric spindle positioning. *PLoS Biol* **7**, e1000088 (2009).
86. Duellberg C CN, Holmes D, Surrey T., C D, NI C, D H, T S. The size of the EB cap determines instantaneous microtubule stability. *eLife* **5**, (2016).
87. D'Alessandro M, *et al.* Amphiphysin 2 Orchestrates Nucleus Positioning and Shape by Linking the Nuclear Envelope to the Actin and Microtubule Cytoskeleton. *Dev Cell* **35**, 186-198 (2015).
88. Shaye DD, Greenwald I. OrthoList: a compendium of C. elegans genes with human orthologs. *PLoS One* **6**, e20085 (2011).
89. Rousselet J, Salome L, Ajdari A, Prost J. Directional motion of brownian particles induced by a periodic asymmetric potential. *Nature* **370**, 446-448 (1994).
90. Preciado Lopez M, *et al.* Actin-microtubule coordination at growing microtubule ends. *Nature communications* **5**, 4778 (2014).

91. Varga V, Helenius J, Tanaka K, Hyman AA, Tanaka TU, Howard J. Yeast kinesin-8 depolymerizes microtubules in a length-dependent manner. *Nat Cell Biol* **8**, 957-962 (2006).
92. Varga V, Leduc C, Bormuth V, Diez S, Howard J. Kinesin-8 motors act cooperatively to mediate length-dependent microtubule depolymerization. *Cell* **138**, 1174-1183 (2009).
93. Von Smoluchowski M. Versuch einer mathematischen Theorie der Koagulationskinetik kolloidaler Lösungen. *Zeitschrift für physikalische Chemie* **92**, 129-168 (1917).
94. Kozlowski C, Srayko M, Nedelec F. Cortical microtubule contacts position the spindle in C. elegans embryos. *Cell* **129**, 499-510 (2007).
95. Bouvrais H, Chesneau L, Pastezeur S, Delattre M, Pecreaux J. Astral microtubule dynamics regulate anaphase oscillation onset and set a robust final position of the C. elegans zygote spindle. *bioRxiv*, (2017).
96. Fink J, *et al.* External forces control mitotic spindle positioning. *Nat Cell Biol* **13**, 771-778 (2011).
97. Thery M, *et al.* Anisotropy of cell adhesive microenvironment governs cell internal organization and orientation of polarity. *Proc Natl Acad Sci U S A* **103**, 19771-19776 (2006).
98. Brenner S. The genetics of Caenorhabditis elegans. *Genetics* **77**, 71-94 (1974).
99. Oegema K, Desai A, Rybina S, Kirkham M, Hyman AA. Functional analysis of kinetochore assembly in Caenorhabditis elegans. *The Journal of Cell Biology* **153**, 1209-1226 (2001).
100. Johnston WL, Krizus A, Dennis JW. Eggshell chitin and chitin-interacting proteins prevent polyspermy in C. elegans. *Curr Biol* **20**, 1932-1937 (2010).
101. Schonegg S, Constantinescu AT, Hoege C, Hyman AA. The Rho GTPase-activating proteins RGA-3 and RGA-4 are required to set the initial size of PAR domains in Caenorhabditis elegans one-cell embryos. *Proc Natl Acad Sci U S A* **104**, 14976-14981 (2007).
102. Redemann S, *et al.* Codon adaptation-based control of protein expression in C. elegans. *Nat Methods* **8**, 250-252 (2011).
103. Consortium CeDM. large-scale screening for targeted knockouts in the Caenorhabditis elegans genome. *G3* **2**, 1415-1425 (2012).
104. Fire A, Xu S, Montgomery MK, Kostas SA, Driver SE, Mello CC. Potent and specific genetic interference by double-stranded RNA in Caenorhabditis elegans. *Nature* **391**, 806-811 (1998).
105. Roul J, Pecreaux J, Tramier M. Procédé de pilotage multi-‐‐modules fonctionnels incluant un dispositif d'imagerie multi-‐‐longueur d'onde, et système de pilotage correspondant Patent EP3123149 A1 (2015).
106. Padilla-Parra S, Auduge N, Coppey-Moisan M, Tramier M. Dual-color fluorescence lifetime correlation spectroscopy to quantify protein-protein interactions in live cell. *Microsc Res Tech* **74**, 788-793 (2011).
107. Cai D, Verhey KJ, Meyhofer E. Tracking single Kinesin molecules in the cytoplasm of mammalian cells. *Biophys J* **92**, 4137-4144 (2007).
108. Pecreaux J, Zimmer C, Olivo-Marin JC. Biophysical active contours for cell tracking I: Tension and bending. In: *2006 Ieee International Conference on Image Processing, Icip 2006, Proceedings*) (2006).

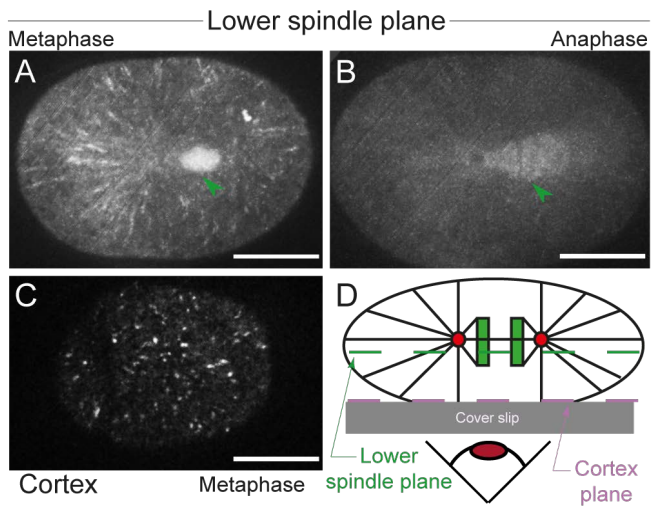

**Figure 1: Dynein intermediate chain DYCI-1::mCherry in the *C. elegans* cytoplasm, on the mitotic spindle, and at the cell cortex.**
Standard deviation map computed over 30 frames from a 5 frames/s DYCI-1::mCherry movie taken in the lower spindle plane (LSP) during (**a**) metaphase and (**b**) anaphase. (**c**) At the cell cortex during metaphase, DYCI-1::mCherry localized in a punctate manner. Green arrowheads indicate the mitotic spindle in the LSP. Scale bars, 10 µm. (**d**) Schematic representation of the spinning disk confocal imaging setup, depicting the spindle through its poles (red disks) from which emanate microtubules (black lines). The chromosomes are the green rectangles. Dashed lines represent the imaging planes, at lower spindle plane (green) and at the cortex (purple).

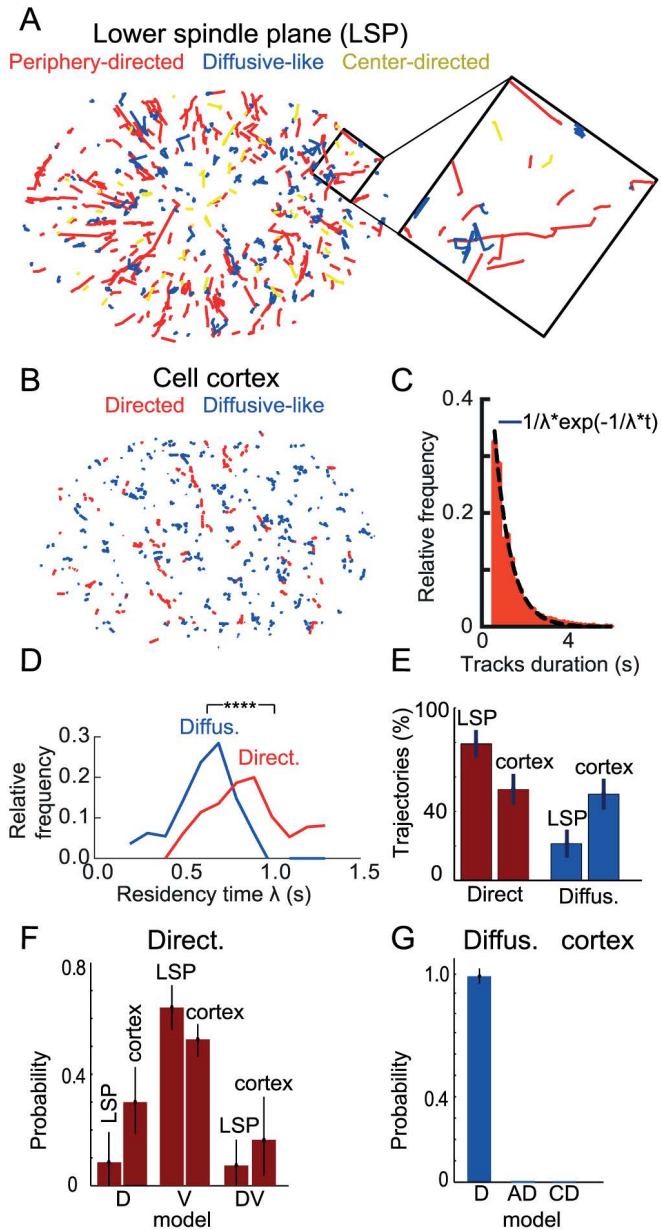

**Figure 2: Dynein intermediate chain DYCI-1::mCherry exhibits different types of motion in the *C. elegans* zygote.**
(**a**) Tracks detected in the lower spindle plane (LSP) divided between those directed towards the cell periphery (red); towards the centre (orange); and those that display a diffusive-like (no clear moving sense) (blue). Inset: zoom highlights the radial alignment of the directed tracks. (**b**) Tracks detected in a movie acquired at the cell cortex with similar color-coding as (**a**). (**c**) Histogram of the track durations at the cell cortex for the diffusive-like tracks of a typical embryo, fitted to an exponential with a residency time λ = 0.7 s. (**d**) Distributions of the residency time λ for directed (red) and diffusive-like (blue) tracks at the cell cortex ($N$ = 26 embryos, 9595 tracks), displaying significantly different mean values ($p ≤ 0.0001$). (**e**) Proportion of directed tracks moving towards the cortex (red) and diffusive-like tracks (blue) in the LSP and at the cortex, averaged respectively over $N$ = 31 embryos (8060 tracks, LSP) and $N$ = 33 embryos (9921 tracks, cortex). (**f**) BCA calculation (Supplementary Text 3.3) of the probabilities of diffusive (D), flow (V), and a mixture of both (DV) models computed on the tracks directed towards the cortex in the LSP and at the cortex averaged over $N$ = 31 embryos (5756 tracks, LSP) and $N$ = 33 embryos (4751 tracks, cortex) respectively. (**g**) Probability of diffusive (D), anomalous super-diffusion (AD), and confined diffusion (CD) models computed on the diffusive-like tracks at the cortex using BCA ($N$ = 33 embryos, 4874 tracks). There were 296 tracks at the cortex that were too short to be classified (see Supplementary Text 3). Error bars indicate the standard error of the mean.

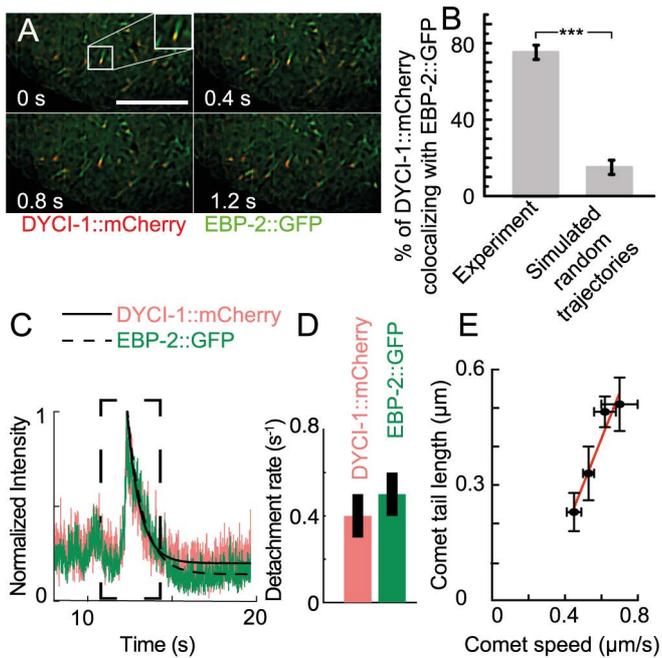

**Figure 3: DYCI-1::mCherry colocalizes with EBP-2::GFP at the microtubule plus-ends.**
(**a**) Sequence of micrographs of the metaphase of a *C. elegans* zygote, showing DYCI-1::mCherry EBP-2::GFP. Scale bar, 10 μm. (**b**) Percentage of DYCI-1::mCherry spots colocalizing with EBP-2::GFP (left, $N$ = 9 embryos, 1880 tracks) compared to those colocalizing with simulated random trajectories (right, Supplementary Text 2.2). Statistical significance was calculated with the Mann-Whitney/Wilcoxon test ($p ≤ 0.001$). (**c**) Intensity profile of a DYCI-1::mCherry and EBP-2::GFP spot crossing the focal volume during a fluorescence correlation spectroscopy experiment and normalized by the peak intensity. Thin black lines show the exponential fits for DYCI-1::mCherry (plain) and EBP-2::GFP (dashed), and the dashed outline indicates the concurrent peaks. (**d**) Detachment rates for $N$ = 8 doubly labelled DYCI-1::mCherry EBP-2::GFP embryos. These were obtained by fitting 43 individual FCS traces, as illustrated in (**c**). (**e**) Linear fit of DYCI-1::mCherry comet tail length (30 to 50 profiles per condition) versus comet speed (typically 7 embryos and 1500 trajectories per condition) for various microtubule growth rates (Fig. S2a). The slope is 1.2 ± 0.2 s, significantly different from zero ($p$ = 0.03) (Supplementary Text 2.3). Error bars indicate the standard error of the mean.

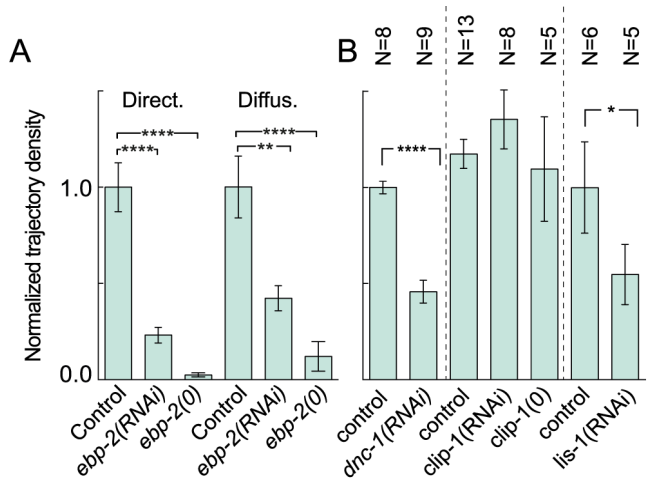

**Figure 4: EBP-2[EB1] +TIP protein contributes to the accumulation of the labelled dynein intermediate chain DYCI-1::mCherry at the microtubule plus-ends in *C. elegans* zygote .**
(**a**) Trajectory densities, divided between directed and diffusive-like motion (Supplementary Text 3.3), normalized by the mean in corresponding control for DYCI-1::mCherry in the lower spindle plane in the control, upon partial *ebp-2(RNAi)*, and after crossing with the *ebp-2(gk756)* null mutation. Data in the LSP are from $N = 8$ control; 11 *ebp-2(RNAi)*; and 5 *ebp-2(gk756)* embryos. (**b**) Similar experiment in control, with results shown upon partial RNA interference in *dnc-1, clip-1,* and *lis-1,* and upon crossing with the *clip-1(gk470)* null mutant. Error bars indicate the standard error of the mean. Stars correspond to the statistical significances (see Methods).

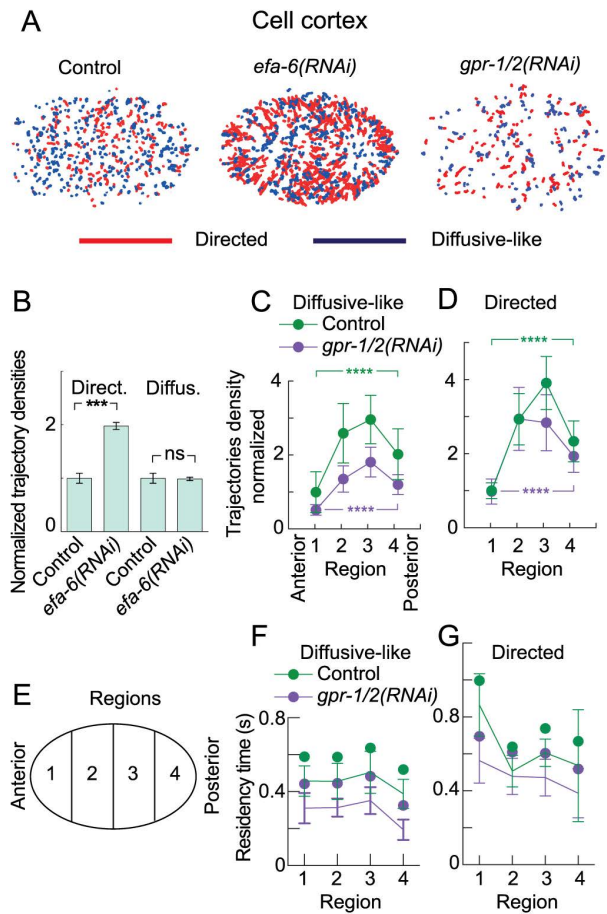

**Figure 5: EFA-6 and GPR-1/2 regulate DYCI-1::mCherry dynamics at the cell cortex.**
(**a**) Tracks detected at the cell cortex in control (left), *efa-6(RNAi)*-treated (middle), and *gpr-1/-2(RNAi)*-treated embryos (right). Tracks were divided between directed (red) and diffusive-like (blue) motion. (**b**) Normalized track densities at the cell cortex in $N = 7$ control embryos (3000 tracks) and $N = 9$ embryos (1400 tracks) treated with *efa-6(RNAi)*. (**c,d**) Diffusive-like and directed track densities at the cell cortex. These were analysed along the AP axis within four regions having equal lengths (see e) in control and *gpr-1/2(RNAi)*-treated DYCI-1::mCherry embryos. Using a Bayesian model approach, we compared raw tracks counts between regions 1 and 4, which are the only relevant to force imbalance (see main text) and found significative differences in all cases (Supplementary Text 2.8). Upon partial *gpr-1/2(RNAi)*, we also observed a reduction of trajectories count by 31% and 15% for anterior- and posterior-tip regions (#1 and #4, respectively) for tracks displaying diffusive-like motion, while only by 10% and 5%, respectively, for the ones with directed motion. (**e**) Schematic of the regions. (**f,g**) Residency times for the embryos plotted in (c,d). We analysed $N = 7$ control (1740 tracks) and $N = 11$ *gpr-1/-2(RNAi)*-treated embryos (2120 tracks). Error bars indicate the standard error of the mean. Stars correspond to the statistical significances (see Methods).

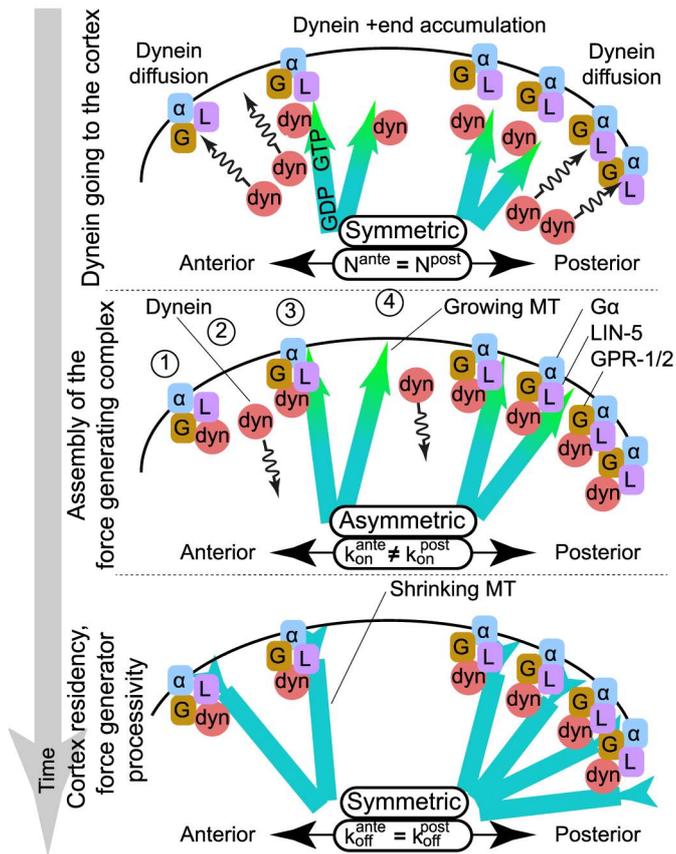

**Figure 6: Arrival and residency of dynein at the cortex highlights the possible causes of force imbalance.**
Schematic views of the arrival, attachment, and residency of dynein at the cell cortex. (**top**) Dynein (red "dyn" disks) arrive at the cortex in equal quantities from the posterior and anterior embryo halves, either by 3D diffusion (squiggly arrows) or after indirect hitchhiking on EBP-2 and accumulation at the plus-ends of growing microtubules (blue/green GTP-capped arrows). Other members of the trimeric complex, GPR-1/2 (orange "G" blobs) and LIN-5 (purple "L" blobs), can be found anchored at the cortex by Gα GPA-16 and GOA-1 (blue "α" blobs). The total amounts of dynein available at the anterior and posterior sides of the cortex are equal. (**middle**) GPR-1/2 (and other complex members) are enriched on the posterior side. They bind the dynein which arrives by 3D diffusion (1) or via plus-end accumulation (3). Dyneins that do not find a cortical anchor (GPR-1/2 complex) leave the cortex (2,4). The attachment rate is therefore higher on the posterior side, which leads to more active force generators in that location. (**bottom**) Bound dyneins are engaged in pulling on astral microtubules that concurrently depolymerize (blue bars). The symmetrical unbinding rate leads to equal residency times in the posterior and anterior sides.

# Movie legends

**Movie S1**. One-cell embryo carrying a randomly integrated DYCI-1::mCherry imaged during metaphase at the lower spindle plane (LSP). Acquisition was done at 5Hz by spinning disk microscopy, with no image processing. The movie was accelerated to 3x real-time. Scale bar, 10 μm.

**Movie S2**. One-cell embryo carrying a randomly integrated DYCI-1::mCherry imaged during anaphase at the lower spindle plane (LSP). Acquisition was done at 5Hz by spinning disk microscopy, with no image processing. The movie was accelerated to 3x real-time. Scale bar, 10 μm.

**Movie S3**. One-cell embryo carrying a randomly integrated DYCI-1::mCherry imaged during metaphase at the cortex. Acquisition was done at 5Hz by spinning disk microscopy, with no image processing. The movie was accelerated to 3x real-time. Scale bar, 10 μm.

**Movie S4**. One-cell embryo carrying a randomly integrated DYCI-1::mCherry and the *dyci-1(tm4732)* null mutation, imaged during metaphase at the lower spindle plane (LSP). Acquisition was done at 5Hz by spinning disk microscopy, with no image processing. The movie was accelerated to 3x real-time. Scale bar, 10 μm.

**Movie S5**. One-cell embryo carrying a randomly integrated DYCI-1::mCherry and the *dyci-1(tm4732)* null mutation, imaged during metaphase at the cortex. Acquisition was done at 5Hz by spinning disk microscopy, with no image processing. The movie was accelerated to 3x real-time. Scale bar, 10 μm.

**Movie S6**. Membrane invaginations pulling from the cytoplasmic membrane towards the cell centre in an embryo doubly labelled by DYCI-1::mCherry (left) and PLCδ1-PH::GFP (centre) upon partial *nmy-2(RNAi)*. Right panel shows the superposition of the left and middle panels. The embryo was imaged at 2.5Hz by spinning disk microscopy. Both channels were filtered by the CANDLE algorithm for better visibility. The movie was accelerated to 7x real-time. Scale bar, 10 μm.

**Movie S7**. Invagination at the cortex of an embryo acquired at 2.5 frame/s in a doubly labelled strain upon *nmy-2(RNAi)*. The green channel is PLCδ1-PH::GFP, and the red is DYCI-1::mCherry. This movie is the source for the stills shown in Fig. S3c.

**Movie S8**. Simulated fluorescence microscopy video depicting diffusive particles with exponential distributed lifetimes, with a time constant of 10 s. The frame rate was 5 Hz, and the signal-to-noise ratio was 5. The movie was accelerated to 3x real-time. Scale bar, 10 µm.

**Movie S9**. Simulated fluorescence microscopy video depicted particles with direct motion with an exponential distributed lifetime, with time constant of 10 s. Frame rate is 5 Hz. Signal to noise ratio was 5. The movie is accelerated to 3x real-time. Scale bar represents 10 µm.

**Movie S10**. Embryo labelled with DYCI-1::mCherry (red) and EBP-2::GFP (green), acquired at the LSP. The movie was filtered by CANDLE and LoG filters (see the Supplementary Text 3.1), and was accelerated to 7x real-time. Scale bar, 10 µm.

**Movie S11**. Embryo labelled with DYCI-1::mCherry (red) and α-tubulin::YFP (green), acquired at the LSP. The inset is a zoom into the region delineated by the white square. The movie is filtered by the CANDLE and LoG filters (see Supplementary Text 3.1) and accelerated to 7x real-time. Scale bar, 10 µm.

# Supplementary figures

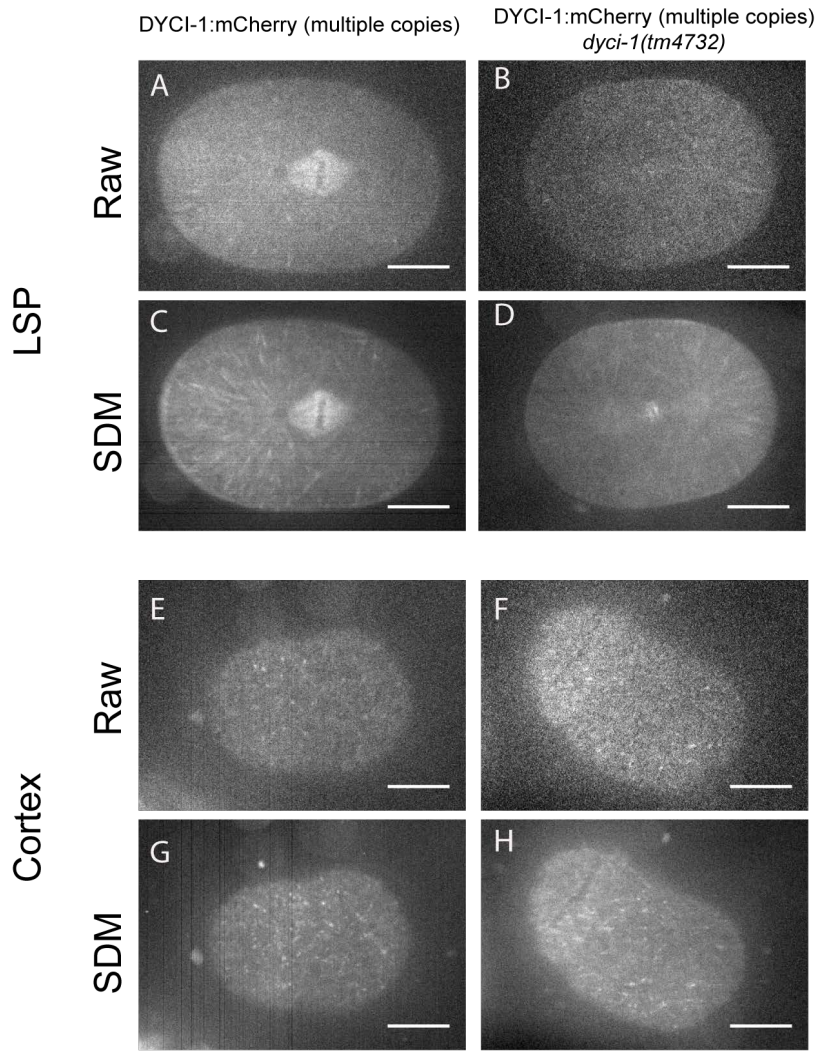

**Figure S1: Micrographs of DYCI-1::mCherry in the lower spindle plane and at the cortex.**
Micrographs of randomly integrated DYCI-1::mCherry, in normal conditions (**a,c,e,g**) and after depletion of endogenous proteins through the *dyci-1(tm4732)* null mutation (**b,d,f,h**). The pictures in (**a,b,e,f**) are raw, while (**c,d,g,h**) are standard deviation maps (SDM, see Materials and Methods) obtained from 125 images acquired at 5 frames/s. Embryos were imaged in the lower spindle plane (LSP) (**a-d**) and at the cortex (**e-h**) by spinning disk microscopy. Scale bars, 10 μm.

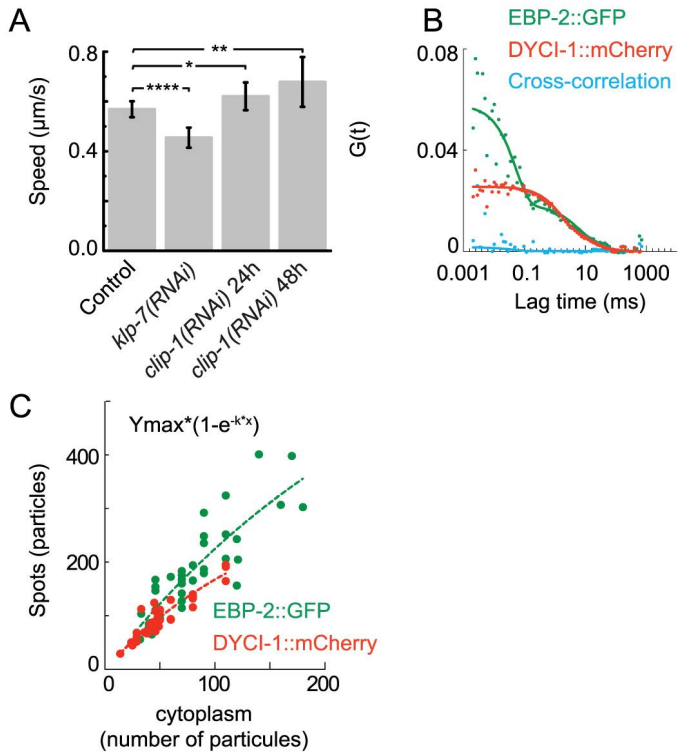

**Figure S2: Dynamics of DYCI-1::mCherry and EBP-2 in the cytoplasm and at the microtubule plus-ends.**
(**a**) Lower spindle plane velocity of DYCI::mCherry spots under four types of modulation of microtubule dynamics. Data come from typically 6 embryos and 1500 trajectories per condition (Supplementary Text 2.3). Error bars indicate standard errors. Stars correspond to the statistical significances (see Methods). (**b**) Autocorrelation of DYCI-1::mCherry (red dots) and EBP-2::GFP (green dots) measured by dual-colour FCS. Experimental curves were fitted to a triplet-state model for each fluorescent species (plain line, same colours). The cross-correlation curve and its fitting using the same model are blue. (**c**) Number of particles in spots compared to the local density of the particles in the cytoplasm, both measured by FCS (see Supplementary Text 1.5 and Fig. 3c for a typical trace) for DYCI-1::mCherry (red) and EBP-2::GFP (green). The dashed lines represent the fit with an exponential growth of the experimental curves based on the equation giving the number of molecules at the plus-end of the microtubule $P_{MT-tip} = Y_{max}\left(1 - e^{-kP_{cyto}}\right)$ as a function of $P_{cyto}$ the cytoplasmic concentration within the FCS focal volume, $k$ is the binding rate and $Y_{max}$ the maximum number of units at the plus-end. This yielded the following fitted values: $Y_{max}^{DYCI-1} = 354$; $k^{DYCI-1} = 0.006$; and $Y_{max}^{EBP-2} = 760$; $k^{EBP-2} = 0.003$ (N = 8 embryos, 43 spots).

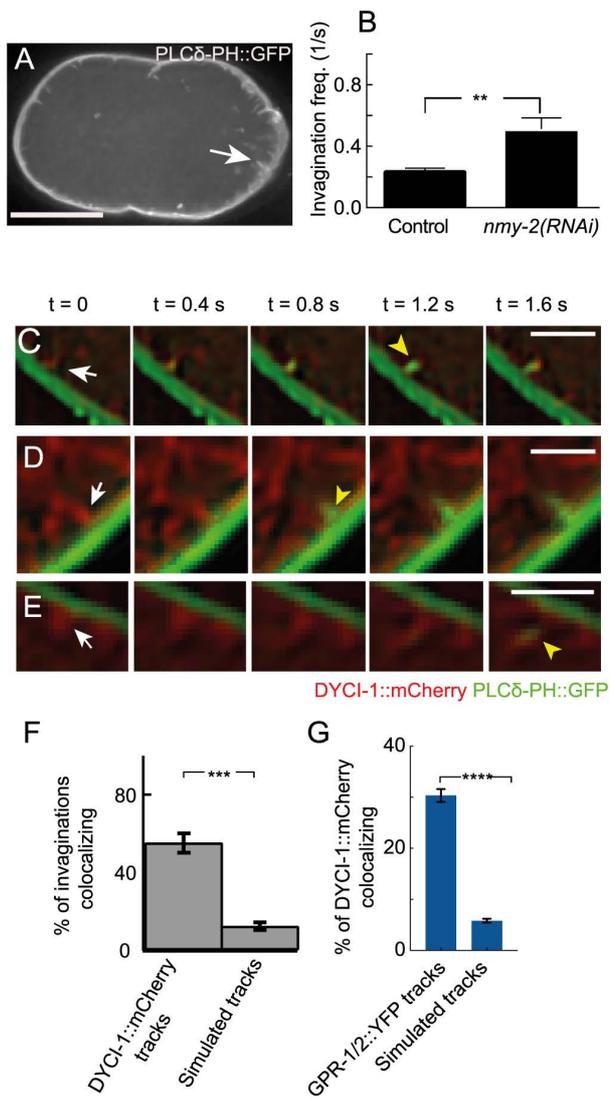

**Figure S3: Dynein spots co-localize with membrane invaginations in the DYCI-1::mCherry PLCδ-PH::GFP strain and with GPR-1/2::GFP.**
(**a**) Maximum intensity projection over 30 frames acquired at 2.5 frames/s for a strain labelled with both DYCI-1::mCherry and PLCδ-PH::GFP, treated by *nmy-2(RNAi)* and viewed in spindle plane (Supplementary text 1.6). Scale bar, 10 µm. The arrow indicates a good example of an invagination. (**b**) Invagination frequencies in the control ($N = 11$ embryos) and in the actin-myosin–cortex–weakened by *nmy-2(RNAi)* embryos ($N = 20$). Error bars indicate standard deviations, and the frequencies are significantly different ($p < 0.001$ per the Mann-Whitney/Wilcoxon test). (**c-e**) Three examples of invagination image sequences acquired during 15 frames at 2.5 frames/s in a doubly labelled strain upon *nmy-2(RNAi)*. Dynein motion and invaginations are viewed from the side. The PLCδ1-PH::GFP channel is green, and DYCI-1::mCherry one is red. One can see dynein arriving at the cortex and, after a brief while, leaving together with an invagination. Membrane invaginations (yellow arrowheads) began after dynein appeared at the cortex (white arrows). Scale bars, 2 µm. (**f**) Percentage of invaginations that colocalize with DYCI-1::mCherry tracks (left, N=18 embryos, 139 invaginations) and those that colocalize with simulated random trajectories in an equal sample (right, Supplementary Text 2.2). (**g**) Percentage of DYCI-1::mCherry tracks that colocalize with GPR-1/2::YFP ones in the doubly labelled strain (left, $N=8$ embryos, 3178 DYCI-1::mCherry tracks and 6373 GPR-1/2::YFP ones) and those that colocalize with simulated random trajectories in an equal sample (right, Supplementary Text 2.2). Statistical significance ($p \leq 0.01$) was calculated using the Mann-Whitney/Wilcoxon test, and error bars indicate the standard error of the mean.

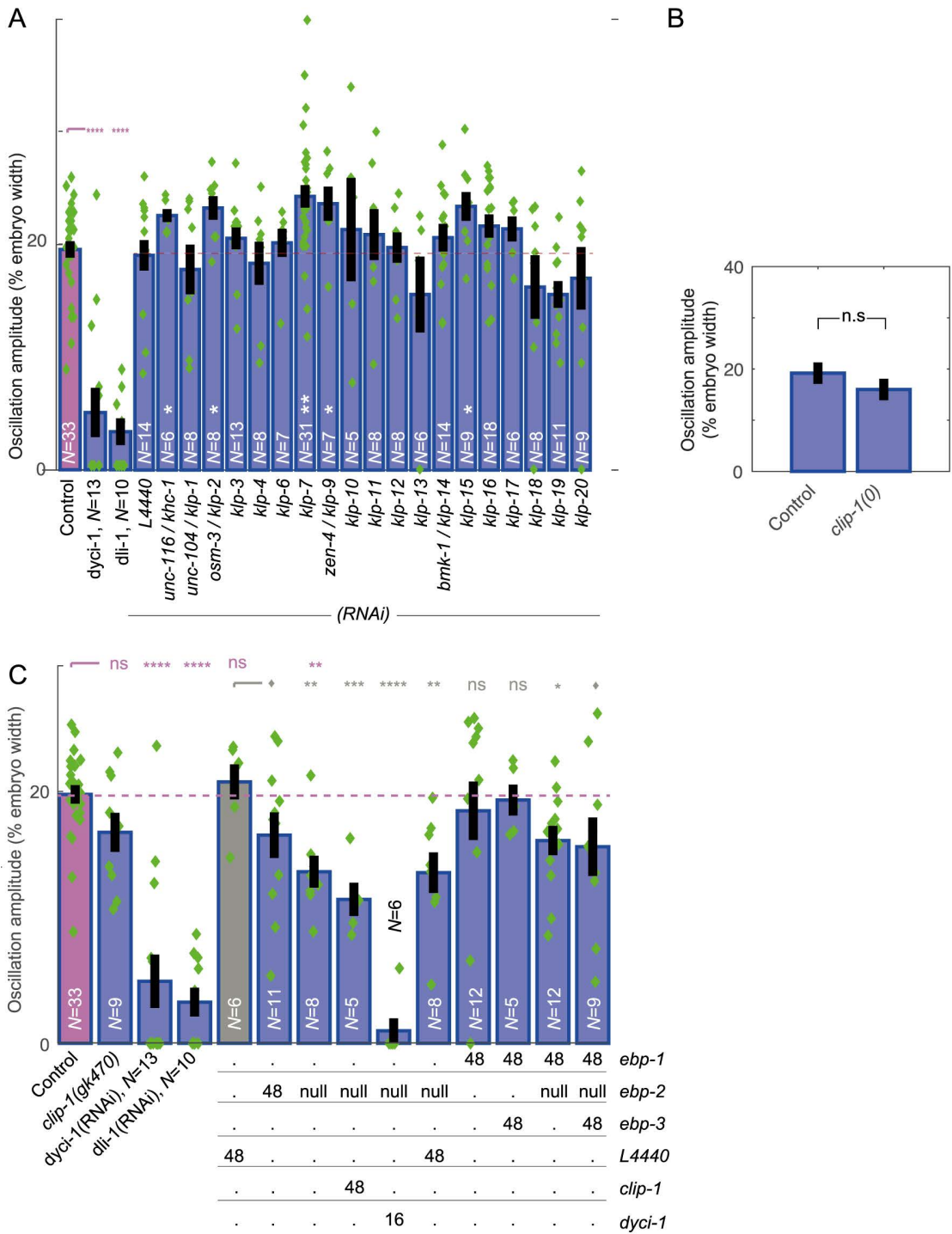

**Figure S4: Posterior centrosomal oscillations upon kinesin, EBP-1/2/3 or CLIP-1 depletion during anaphase.**
(**a**) Maximum amplitudes as a percentage of embryo width for anaphase posterior centrosome oscillations. Values were obtained upon partial RNAi of each kinesin, with DYCI-1 and DLI-1 partial depletions also listed for reference. Green diamonds are raw data, and the horizontal dashed pink line indicates the amplitude for non-treated embryos. Pink stars signal a significant differences compared to the control, and white ones compared to *L4440(RNAi)*. (**b**) Posterior centrosome oscillation maximum amplitudes as percentages of embryo width in $N$ = 10 control γ-tubulin TBG-1::GFP embryos and $N$ = 10 embryos crossed with the *clip-1(gk470)* null mutant. Results were not significant. (**c**) Maximum amplitudes as percentages of embryo width in anaphase posterior centrosome oscillations upon various depletions of the EB1 homologs EBP-1/2/3, with DYCI-1, DLI-1 and CLIP-1 partial depletions shown for reference. The table below the plot indicates the conditions null for the *ebp-2(gk756)* null mutant; and 48 or 16, which is the number of feeding hours for RNAi experiments. Green diamonds correspond to raw data, and the horizontal pink dashed line indicates the amplitude for non-treated embryos. Pink stars show the significance with respect to non-treated embryos, and grey ones with respect to control RNAi with the L4440 vector. Error bars correspond to standard errors of the means.

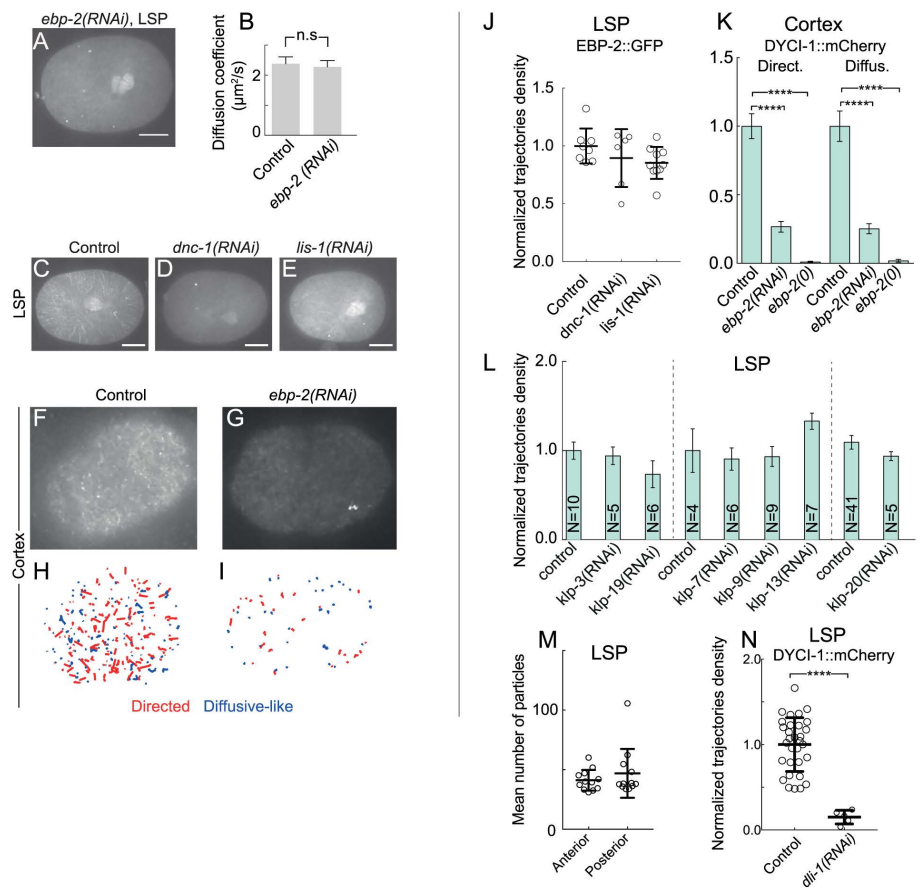

**Figure S5: DYCI-1::mCherry directed motion in the LSP is dependent on EBP-2, dynactin, and LIS-1, while cortex localization depends on EBP-2.**
(**a**) Maximum intensity projection computed from a 100-frame DYCI-1::mCherry movie (5 frames/s) upon partial *ebp-2(RNAi)*. (**b**) DYCI-1::mCherry diffusion coefficients measured by FCS in the cytoplasm for the doubly labelled DYCI-1::mCherry EBP-2::GFP strain in control and upon *ebp-2(RNAi)* ($N$ = 5 embryos, 3 FCS spots/embryo). Difference was not significant (n.s.) Error bars are standard errors of the mean. (**c-e**) Maximum intensity projection computed from a 100-frame movie (5 frames/s) of (**c**) untreated DYCI-1::mCherry; (**d**) upon *dnc-1(RNAi)* treatment; and (**e**) upon *lis-1(RNAi)* treatment. Scale bars, 10 µm. Typical (**f**) control and (**g**) *ebp-2(RNAi)*-treated DYCI-1::mCherry embryos imaged at the cortex after CANDLE pre-processing to enhance visibility (Supplementary Text 3.1). (**h,i**) Tracking as per (**f,g**), respectively, with directed and diffusive-like motions shown. (**j**) Trajectory densities at the cortex normalized by the EBP-2::GFP control mean in control embryos ($N$ = 8); *dnc-1(RNAi)* dynactin subunit-treated embryos ($N$ = 6); and *lis-1(RNAi)* ones ($N$ = 10). Circles denote individual embryo values. (**k**) Same as (**j**) but in the cell cortex and normalized against the DYCI-1::mCherry control ($N$ = 6); upon partial *ebp-2(RNAi)* ($N$ = 9); and after crossing with the *ebp-2(gk756)* null mutation ($N$ = 3). Differences are highly significant. (**l**) Normalized trajectory densities of DYCI-1::mCherry in the LSP in control and upon partial RNAi of selected kinesins. No significant differences were found in the kinesins. (**m**) Number of particles in the anterior and posterior embryo halves as measured by FCS. Each circle corresponds to a single embryo. (**n**) Trajectory densities at the cortex normalized by the DYCI-1::mCherry control mean in control embryos ($N$ = 31); *dli-1(RNAi)* dynein light intermediate chain subunit-treated embryos ($N$ = 5). Circles denote individual embryo values. Error bars indicate the standard errors of the means (**k,l**) or standard deviations (**j, m, n**).

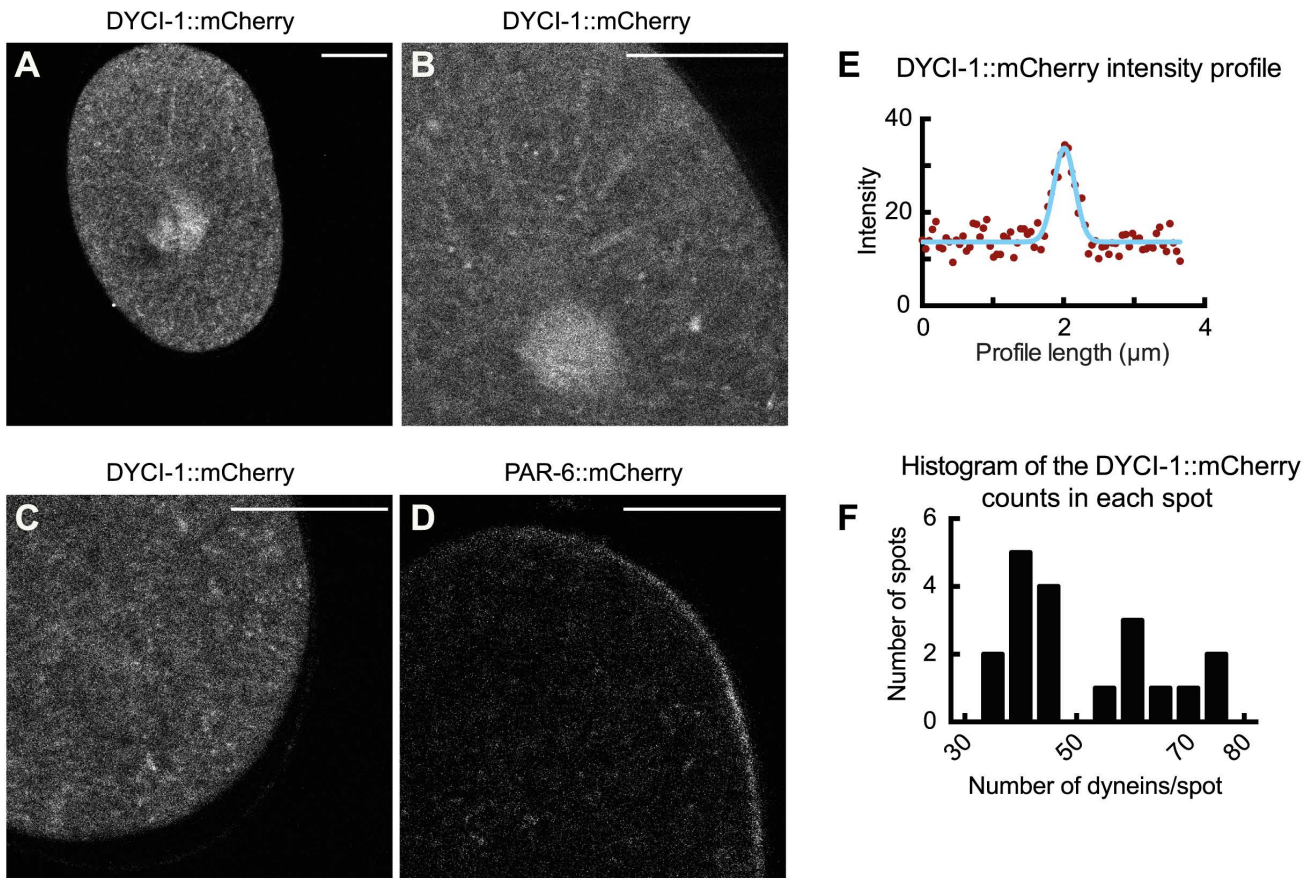

**Figure S6: Using image density to count the number of dynein per spot at the microtubule plus-ends.**
Typical micrographs of (**a-c**) randomly integrated DYCI-1::mCherry used to measure spot intensity profiles and (**d**) PAR-6::mCherry used to calibrate intensity in a number of mCherry dyes. Scale bars, 10 µm. (**e**) Exemplary fit of a DYCI-1::mCherry spot intensity profile by a Gaussian distribution with background. (**f**) Histogram of the intensities of the spots measured by the amplitude obtained through the fitted Gaussian distribution after the background has been subtracted.

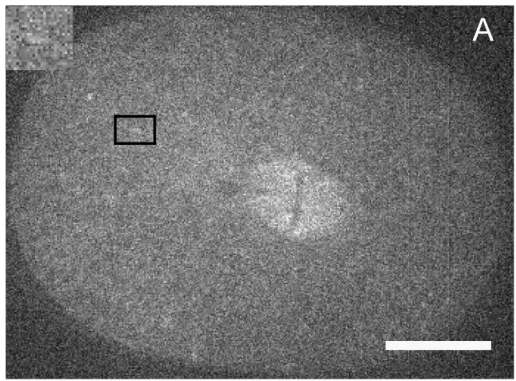

Raw image (SNR = 3)

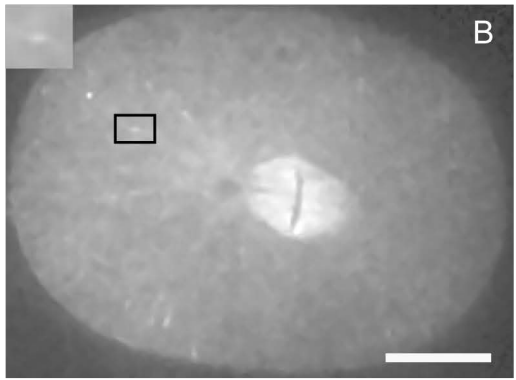

CANDLE filtering (SNR = 7)

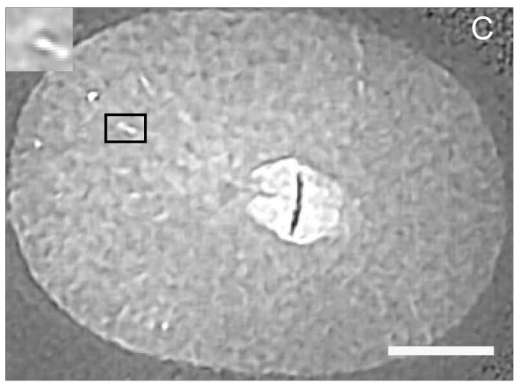

LoG filtering (SNR = 12)

**Figure S7: A typical DYCI-1::mCherry micrograph analysed using our tracking pipeline.**
One-cell embryo during metaphase acquired after an exposure of 0.2 s, with signal-to-noise ratios indicated. Insets are magnified views of the region delineated by a black rectangle and highlighting a spot. Scale bars, 10 μm. The images were obtained at each step of pre-processing: (**a**) raw picture; (**b**) image denoised by the CANDLE filter; (**c**) image after 2-D LoG filtering (see Supplementary Text 3.1).

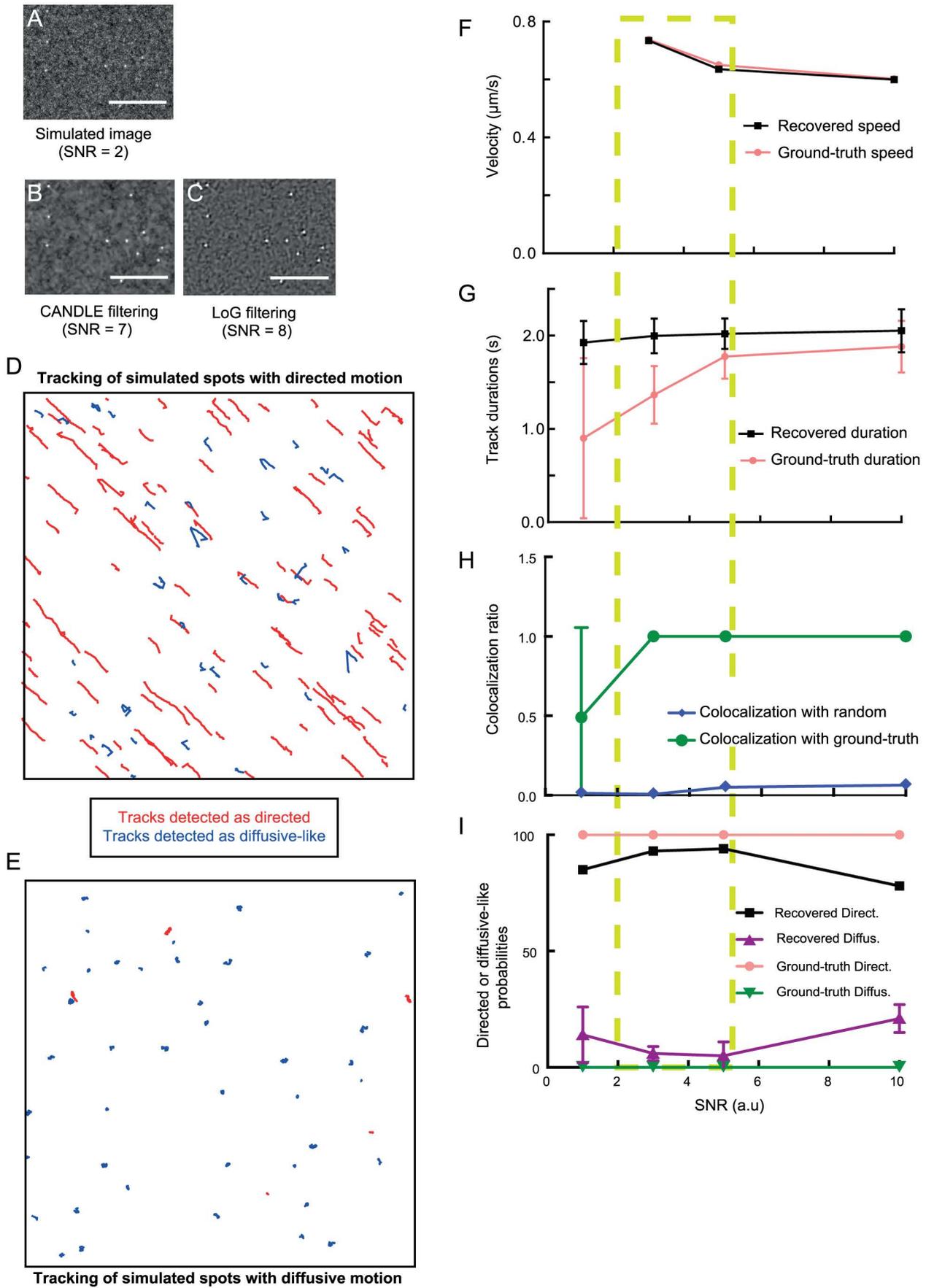

**Figure S8: Validating our tracking pipeline by using fabricated images that mimic experimental ones.**
Fabricated images with known spot dynamics and mimicking experimental ones (see Supplementary Text 3.4) analysed with our pipeline. The signal-to-noise ratios (SNRs) are indicated, and the scale bars are 10 μm. The images were obtained at each step of pre-processing: (**a**) raw picture; (**b**) image denoised by the CANDLE filter; (**C**) image after 2-D LoG filtering (see Supplementary Text 3.1). (**d,e**) Typical tracking of two movies in which simulated spots displayed directed (**d**) and diffusive (**e**) motions. The tracks recovered by our analysis pipeline are shown in red for directed motion and in blue for diffusive-like. (**f-g,i**) The light green dashed rectangle highlights the SNR values that correspond to our experimental observations. From directed tracks, we compared (**f**) average track speeds and (**g**) durations from the simulation (ground-truth, pink circles) and from our analysis (black squares). (**h**) We also tested the colocalization of the tracks (green circles) compared to random colocalizations (blue diamonds). (**i**) We then challenged the classification by computing the recovered ratio percentages of the tracks classified as directed (black squares) or diffusive-like (upward-pointing purple triangles) and compared these to the set values in the simulation for directed (pink circles) and diffusive-like ones (downward-pointing green triangles). Error bars indicate the standard error of the mean.

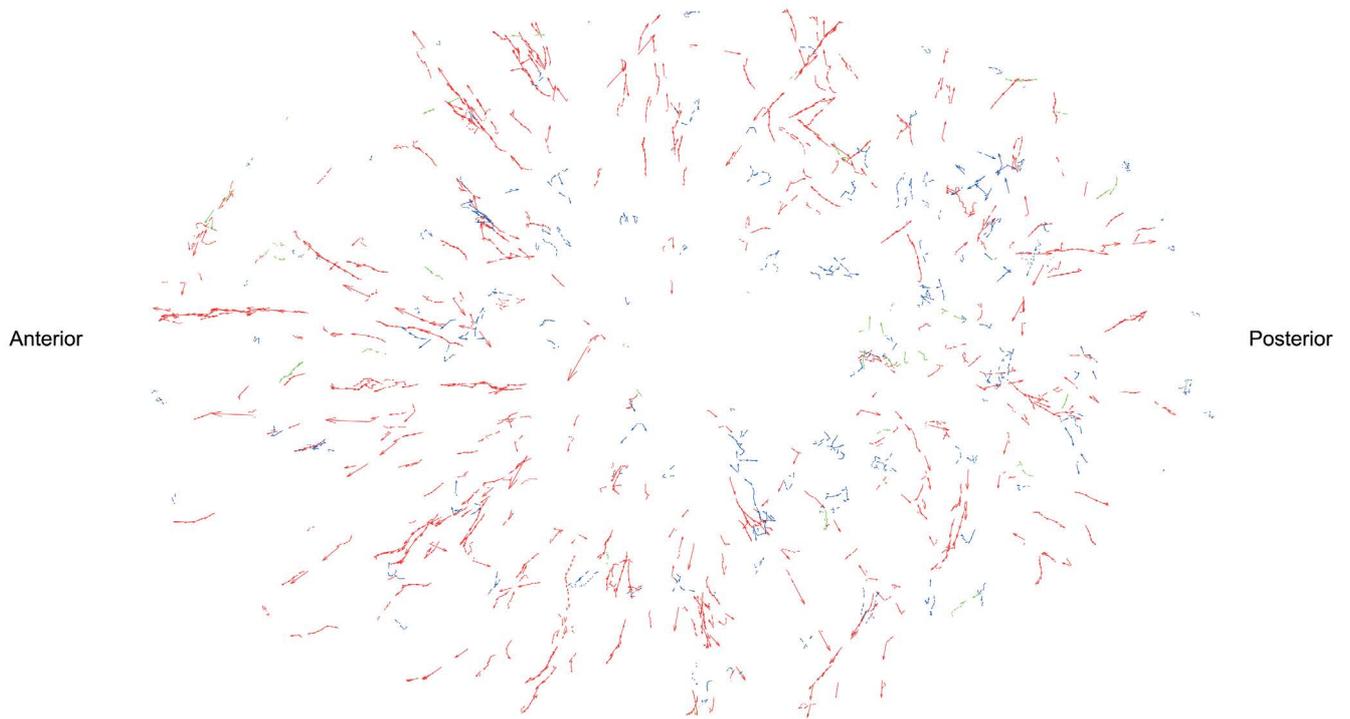

**Figure S9: DYCI-1::mCherry velocity map of spots acquired at the LSP.**
Velocity map obtained in the LSP for the same embryo as shown in Fig. 1a. DYCI-1::mCherry spots are depicted by small arrows having lengths proportional to their instantaneous speed. The trajectories were classified between those directed at the cell periphery (red), towards the centre (green), or having diffusive-like motion (blue).

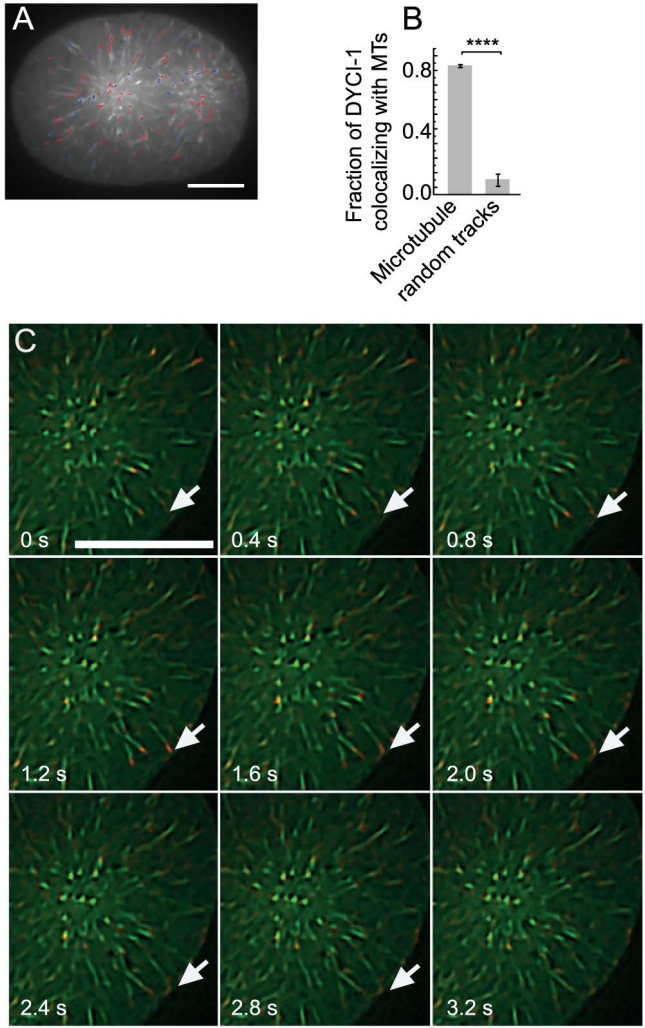

**Figure S10: Directed tracks of DYCI-1::mCherry spots colocalize with microtubules.**
(**a**) DYCI-1::mCherry tracks, divided into those directed to the cell periphery (red), towards the centre (orange), and diffusive-like motion (blue), are superimposed on the maximum intensity projection of microtubules (α-tubulin TBA-2::YFP) computed from a 10-frame dual-colour stack acquired at 2.5 frames/s. Scale bar, 10 µm. (**b**) Fraction of DYCI-1::mCherry spots colocalizing with microtubules as compared to those colocalizing with simulated random tracks (Supplementary Text 2.2; *N*=5 embryos, 9 stacks, 200 tracks). Statistical significance ($p < 10^{-4}$) was computed with the Mann-Whitney/Wilcoxon test, and error bars show standard errors. (**c**) Sequence of micrographs acquired every 0.4 s, displaying DYCI-1::mCherry (red) and α-tubulin::YFP (green). Arrows indicate microtubule tips where DYCI-1::mCherry accumulates. Scale bar, 5 µm.

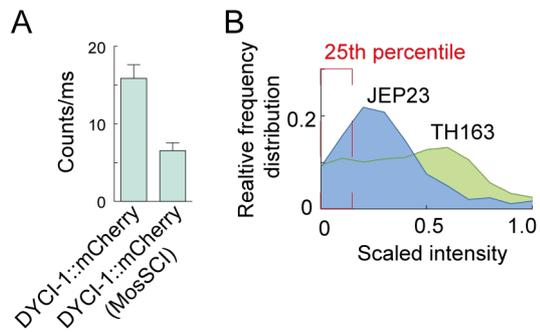

**Figure S11: DYCI-1::mCherry expression levels and detection threshold.**
(**a**) The average intensities (counts/ms) in the cytoplasm obtained by FCS for TH163, the randomly integrated DYCI-1::mCherry ($N$ = 6 embryos, 38 spots), and JEP23, carrying exactly two copies of DYCI-1::mCherry ($N$ = 10 embryos, 52 spots). Error bars represent the standard errors of the mean. (**b**) Intensity histogram of the tracked spots in TH163 (green area, $N$ = 6 embryos, 59 spots) and JEP23 (blue area, $N$ = 6 embryos, 52 spots). The intensity values were scaled from 0 to 1 for plotting, with 0 being the minimum value and 1 the maximum. The dashed red line represents the 25 percentile for JEP23.

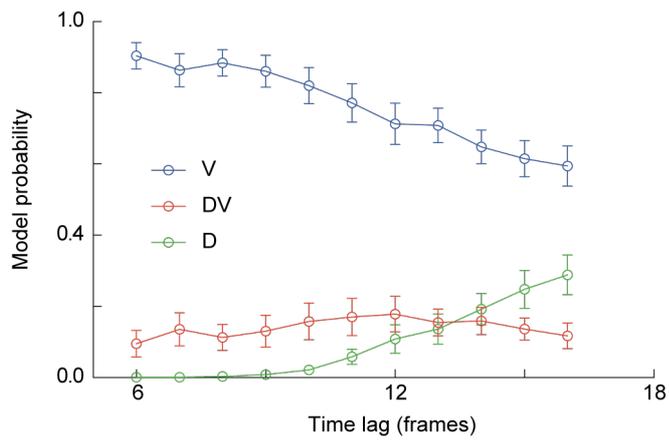

**Figure S12: Typical probabilities of concurrent models in Bayesian classification analysis (BCA) upon varying the mean square displacement time lag.**
Mean probability values for 3 models (V, DV, D) using the BCA classification pipeline (see Supplementary Text 3.3) and computed over $N = 31$ embryos. The average values are shown against the different lag times used to calculate the mean square displacement. The error bars represent the standard error of the mean between embryos for each lag time. The graph corresponds to an archetypical case of directed tracks in control embryos in the LSP.

# Supplementary Text to "Dynein dynamics at the microtubule plus-ends and cortex during division in the *C. elegans* zygote"

Ruddi Rodriguez-Garcia[1,2,3,*], Laurent Chesneau[1,2,*], Sylvain Pastezeur[1,2], Julien Roul[1,2,4], Marc Tramier[1,2], & Jacques Pécréaux[1,2,#]

[1] CNRS, UMR6290, Institute of Genetics and Development of Rennes (IGDR), 35043 Rennes, France. [2] University of Rennes 1, UEB, SFR Biosit, School of Medicine, 35043 Rennes, France. [3] Present address: Cell Biology, Faculty of Science, Utrecht University, Padualaan 8, 3584 CH Utrecht, The Netherlands. [4] Present address: LAAS - Laboratoire d'analyse et d'architecture des systèmes, 7 avenue du Colonel Roche, BP 54200, 31031 Toulouse cedex 4, France.

\* These authors contributed equally to this work.

\# Correspondence and requests for materials should be addressed to J.P. (email: jacques.pecreaux@univ-rennes1.fr)

Running title: Dynamics of dynein in the *C. elegans* zygote

# 1 DYCI-1::MCHERRY IS A BONA FIDE REPORTER OF DYNEIN DYNAMICS.

In our paper, we set out to offer a detailed view of the dynamics of dynein at the microtubule plus-ends, clarifying how it is efficiently targeted to the cell periphery and whether it is a transport mechanism, and measuring its activity at the cell cortex. To achieve this, we used a dynein light chain fluorescent reporter (Fig. 1 and Movies S1-3), then did tracking and a detailed quantification (Fig. 2). To take full advantage of quantification, one must ensure that the fluorescent transgene faithfully reports all dynein localizations, without artefacts. It is equally important that the labelled and endogenous dyneins perform similarly, in particular in their contributions to the pulling forces generated at the cell cortex.

## 1.1 DYCI-1::MCHERRY IS A FUNCTIONAL TRANSGENE

To investigate the *in vivo* dynamics of dynein targeting to the cortex and its residency there, we used the TH163 strain carrying a randomly integrated transgene which encodes the fluorescently labelled DYCI-1::mCherry dynein sub-unit. This transgene was flanked with the endogenous regulatory sequences and produced within the framework of the TransgeneOme project (Sarov et al., 2012; Sarov et al., 2006). Fluorescence intensity in the one-cell embryo showed variability in the expression, but there was no significant phenotype or lethality when compared to wild type (N2, $N > 30$). In *C. elegans* zygote, in contrast to mammalian cells, DYCI-1 is the only intermediate chain of dynein needed in early embryos, while its PANTHER-identified paralog C27F2.1 (Mi et al., 2010; Mi et al., 2013), has no early embryo phenotype (Kamath et al., 2003; Sonnichsen et al., 2005). We can therefore expect this transgene to report dynein in all its functions.

## 1.2 DYCI-1::MCHERRY TRANSGENE RESCUE OF RECESSIVE EMBRYONIC LETHAL NULL MUTATION *DYCI-1(TM4732)*.

We first aimed to validate the functioning of the randomly integrated transgene by rescuing the *dyci-1(tm4732)* lethal recessive null mutation using the transgene. We began by producing JEP9. This was done by crossing the heterozygous strain carrying the DYCI-1 *tm4732* null mutation with the VC2542 strain. This latter carries a balancer featuring a pharynx/gutGFP marker and the re-arrangement nT1[qIs51] which prevents hermaphrodite to have progeny (and two wild-type *dyci-1* copies). The JEP9 strain showed a lethality rate of 48.8±7.9% ($N$ = 3 replica, 302 hermaphrodites), which suggested that homozygous *dyci-1(tm4732)* must be 100% lethal. This was confirmed by counting in JEP9 progeny, the GFP negative larvae, i.e. homozygous for *dyci-1(tm4732)*. We found a frequency of 0.3±0.3% (N=3 replica, 498 hermaphrodites).

We then obtained JEP40 by crossing JEP9 twice with TH163, a randomly integrated DYCI-1::mCherry strain. JEP40 has only mCherry-labelled *dyci-1* genes, without endogenous copy. It displayed a reduced embryonic lethality, 20.5±7% ($N$ = 3 replica, 123 hermaphrodites), and some dpy/unc phenotypes. We noticed a variation in fluorescence levels in the randomly integrated DYCI-1::mCherry, therefore we decided to explore expression levels. With the help of the Biology of *Caenorhabditis elegans* facility (UMS3421, Lyon, France), we used MosSCI to design JEP23, a strain carrying exactly two copies of this transgene (Boulin and Bessereau, 2007; Robert and Bessereau, 2007). For this purpose, the DYCI-1::mCherry sequence (consisting of 3035 nucleotides from before the dyci-1 5'UTR to the end of the 3'UTR) was amplified from TH163 genomic DNA, cloned into the Afl II/BglII site of the pCFJ352 plasmid, and bombarded into the EG6701 strain. Integration on the ttTi4348 MosSCI site was verified by sequencing. We twice crossed this JEP23 strain with the *dyci-1(tm4732)* null mutation JEP9 to produce JEP30. Worms homozygous for dyci-1::mCherry insertion and heterozygous for the *dyci-1(tm4732)* null mutation produced viable null mutant homozygote worms, whose progeny displays 81.1±14.2% of lethality (*N* = 3 replica, 21 hermaphrodites). We concluded that the DYCI-1::mCherry transgene products can perform all of the roles of DYCI-1.

We then focused on the phenotypes of the progeny of homozygous *dyci-1(tm4732)* strains at the one-cell stage. JEP40 strains also carrying randomly integrated DYCI-1::mCherry, had no particular phenotype. In contrast, in the second generation of the JEP30 strain, obtained by crossing JEP23 with JEP9 and having exactly two copies of dyci-1::mCherry, we observed a phenotype at the one-cell stage that is similar to that of *dyci-1(RNAi)* when observed between 12 and 19 h after transfer onto feeding plates. Going further, we rarely observed more than 2 pronuclei (2/11 embryos). In embryos with two pronuclei, meeting happened in 6/9 embryos. In these, the spindle was often ill-formed (5/6) but still usually at least partially migrated to the cell centre and oriented along the antero-posterior axis (4/6) prior to the onset of cytokinesis. Initial furrow ingression was well-positioned (4/6). Overall, we concluded that the one-cell stage *dyci-1(tm4732)* phenotype is rescued by the dyci-1::mCherry construct, although the expression level of DYCI-1::mCherry transgene, when present in exactly two copies, might be too low.

## 1.3  DYCI-1::MCHERRY FLUORESCENT SPOTS ARE NOT ARTEFACTS OF OVEREXPRESSION.

Because the number of DYCI-1::mCherry transgene copies is not constrained in the strain in which it was randomly integrated, maybe the DYCI-1 spots are non-physiological, due to overexpression. We imaged the strains devoid of endogenous *dyci-1* (Fig. S1). As with the strains having exactly two copies, we were able to successfully detect spots in the LSP and at the cortex in the randomly integrated DYCI-1::mCherry (Fig. S1b-d, f-h, and Movies S4-5). However, in the case of exactly two copies, the spots were much fainter, making quantification impossible. We concluded that the spots observed in the randomly integrated DYCI-1::mCherry strain carrying endogenous copies (referred to hereafter simply as "DYCI-1::mCherry") are not artefactual, and are representative of dynein dynamics.

## 1.4  MEASURING CYTOPLASMIC CONCENTRATIONS BY FCS

To understand the dynamics of dynein at the cell cortex and its ties to force generation, we had to test for a potential asymmetry in dynein concentrations. Furthermore, in order to validate our dynein labelling, we also needed to demonstrate that the cytoplasmic dynein spots are dynamic (see 1.1.5), which requires knowing their cytoplasmic concentrations. To do so, fluorescence correlation spectroscopy (FCS) traces (Fig. 3c) were analysed using SymPhoTime (PicoQuant). We fitted the autocorrelation function $G(\tau)$ using a triplet-state model for one fluorescent species (Widengren et al., 1994) as follows:

$$G(\tau) = G(0) \left[ 1 - T + Te^{\left(-\frac{t}{\tau_T}\right)} \right] \left(1 + \frac{t}{\tau}\right)^{-1} \left(1 + \frac{t}{\tau \kappa^2}\right)^{-1/2} \qquad (S1)$$

where *t* is the lag time in ms; *T* is the triplet decay fraction; $\tau_T$ is the lifetime of the triplet state in ms; $\tau$ is the diffusion time of the fluorescent species in ms; and $\kappa$ is the length-to-diameter ratio of the focal volume set to 4. We calculated the cytoplasmic concentration as $C = \frac{1/G(0)}{V_{eff} N_A}$, where $V_{eff}$ is the effective excitation volume and $N_A$ is the Avogadro constant. We found $32 \pm 11$ particles in the focal volume, estimated at 0.3 fl, leading to an estimated concentration of $177 \pm 60$ nM.

## 1.5 MEASURING DYNEIN BINDING RATES AT MICROTUBULE PLUS-END.

To analyse the dynamics of dynein spots and estimate the binding rate of dynein to a microtubule plus-end, we followed (Dragestein, 2008) and fitted the number of units bound at the plus-end versus the concentration in the cytosol nearby into the equation:

$$P_{MT-tip} = Y_{max}\left(1 - e^{-k P_{cytosol}}\right) \qquad (S2)$$

where $P_{MT-tip}$ is the number of DYCI-1::mCherry at the plus-end of the microtubule; $P_{cytosol}$ is the cytoplasmic concentration within the FCS focal volume; $k$ is the binding rate; and $Y_{max}$ is the maximum number of units at the plus-end. We measured the cytoplasmic concentration as detailed above. The number of particles in the peak was obtained by multiplying the ratio of peak-to-basal intensity in the FCS trace (e.g. Fig. 3c) by the corresponding cytoplasmic concentration expressed as a number of particles. This assumes that the basal intensity corresponds to the cytoplasmic concentration. We analysed 43 spots on 8 embryos of the double-labelled strain expressing DYCI-1::mCherry and EBP-2::GFP. For each channel, we used equation S2 to fit the respective measurements. DYCI-1::mCherry and EBP-2::GFP displayed similar dynamics, in particular having close binding rates of $k^{DYCI-1} = 0.006$ and $k^{EBP-2} = 0.003$ (Fig. S2c). Using fluorescence cross correlation spectroscopy (FCCS), we showed that these two molecules are not associated in the cytoplasm (Fig. S2b). Besides teaching us about dynein dynamics at the microtubule plus-ends, this also indicates that the spots are relevant to dynein dynamics, that they are not stable or artefactual aggregates.

## 1.6 TUBE ASSAY IN THE STRAIN DOUBLY LABELLED WITH MEMBRANE AND DYNEIN.

On the functional side, we were interested in dynein's contribution to cortical pulling force generation (Nguyen-Ngoc et al., 2007; Pecreaux et al., 2006). We tested whether cortical DYCI-1::mCherry is involved in this using the previously described "tube assay"(Redemann et al., 2010). Partial *nmy-2(RNAi)* was first used to weaken the cortex while preserving polarity. Once the actin-myosin cortex was weakened, the positions of the cortical force generators were revealed by their pulling of the cytoplasmic membrane tubes towards the centrosomes. We started by generating a new strain (JEP20) expressing both DYCI-1::mCherry and a labelled PH domain of phospholipase PLCδ1-PH::GFP, to enable visualization of the plasma membrane (Fig. S3a and Movie S6). We found that upon *nmy-2(RNAi)*, $55 \pm 5\%$ of the invaginations displayed DYCI-1::mCherry labelling ($N = 18$ embryos, 139 invaginations; see the main text, Fig. S3c-e, and Movie S7).

Interestingly, without treatment we observed $22 \pm 13$ invaginations ($N = 20$ embryos), with a frequency occurrence of $0.51 \pm 0.07$ s$^{-1}$ during anaphase. This compares to $9 \pm 7$ invaginations and a frequency of $0.23 \pm 0.03$ s$^{-1}$ in the control strain with the PLCδ1-PH::GFP labelling only (*N* = 11 embryos: Fig. S3b). The invagination count obtained in non-treated DYCI-1::mCherry strain is also larger than previously reported for non-treated embryos (Redemann et al., 2010). More importantly, we observed that 42 ± 7% of membrane invaginations are dynein-decorated in these not treated embryos (*N* = 8 embryos, 84 invaginations). In conclusion, because of the good colocalization of dynein and invaginations, with or without a weakened actomyosin cortex, we suggest that labelled dynein may be involved in cortical pulling force generation.

Why were half of the invaginations not visibly tagged by DYCI-1::mCherry? It is likely that they contained an amount of DYCI-1::mCherry that fell below our detection limits. The physics of invaginations were studied *in vitro,* and force in the order of tens of pN could be sufficient to pull an invagination (Dernyi et al., 2002; Leduc et al., 2004). The stall force for a dynein is estimated at about 6 pN (Howard, 2001). We estimated our detection threshold (see 1.1.7 below) to be 26 dyneins, which is much higher than the number of dyneins required to produce a threshold force to pull an invagination. It is therefore probable that some invaginations had dynein counts below this threshold. We concluded that DYCI-1::mCherry correctly reports the dyneins relevant to cortical force generation, and it is therefore appropriate to investigate the dynein localizations and dynamics using this strain.

## 1.7 ESTIMATING THE DETECTION THRESHOLD.

We figured that the large cytoplasmic concentration of dynein might limit our ability to detect spots in made of a low number of DYCI-1::mCherry. Since it is close to our detection limit (see the cytoplasmic intensities in Fig. S11a), we computed the detection threshold by considering the microtubule plus-end spot intensities in JEP23, the strain with exactly two copies of DYCI-1::mCherry integrated by MosSCI. We assumed that the spot intensities were distributed normally above a constant background. We fitted the histogram of spot intensities using a Gaussian distribution with an added constant, for both JEP23 (*N* = 6 embryos, 41 spots) and the TH163 randomly integrated DYCI-1::mCherry-carrying strain (*N* = 6 embryos, 52 spots). The added constant values corresponded to the background levels. We next subtracted the respective background levels from the spot intensity distribution for each strain, and computed the 25$^{th}$-percentile (Fig. S11b), using the result as the intensity-detection threshold.

To convert this value into a particle count, we assumed that the average of the intensity distribution, having subtracted the background, for the doubly labelled DYCI-1::mCherry/EBP-2::GFP strain is the same as that of JEP23. In other words, the previously described P$_{MT-tip}$ results were equal for both strains. We could thus associate a number of particles to the average DYCI-1::mCherry spot intensity, and therefore to the 25$^{th}$-percentile of JEP23 intensity distribution. In this case, we ended up with a threshold of 26 ± 4 particles.

## 1.8 DYCI-1::MCHERRY REMAINS ASSOCIATED TO THE DYNEIN COMPLEX

To solidify our approach, we sought indications that DYCI-1 remains associated to the dynein complex in the cytoplasm. This would indicate that association would be also true at the microtubule plus-ends since dynein complex binds there coming from the cytoplasm. To do so, we used FCS to estimate the size of labelled "particles" in the cytoplasm and *in vivo* by measuring their diffusion coefficients *D*. We used PAR-6::mCherry as a control for which we measured $D_{par-6}$ = 15 ± 3 nm$^2$s$^{-1}$ (*N* = 4 embryos, 12 spots) and then we found *D*

= 2.6 ± 0.7 nm²s⁻¹ ($N$ = 9 embryos, 38 spots) and. To interpret these results, we inferred viscosity of the cytoplasm from the PAR-6::mCherry control, and used the Stokes-Einstein-Sutherland equation to compute the hydrodynamic radius for the complex associated with DYCI-1::mCherry. From what is known about its domains (Garrard et al., 2003; Hirano et al., 2005) as identified by InterPro (Mitchell et al., 2015), we assumed that PAR-6 is globular-shaped. Further, from the number of residues in PAR-6::mCherry transgene, we estimated its radius to be $r_{PAR-6}$ = 2.9 ± 0.8 nm using (Wilkins et al., 1999) and propagating the errors in the formula to assess uncertainty. This enabled us to estimate the viscosity using the PAR-6::mCherry diffusion coefficient. Combining this viscosity and the DYCI-1::mCherry diffusion measurement, the Stokes-Einstein-Sutherland equation yields $r_{DYCI-1}$ ≈ 17.0 ± 7.3 nm. This corresponds to the value measured for the human dynein dimer (Trokter et al., 2012), which also has an estimated molecular weight (723 kD) comparable to that of *C. elegans* (675kD). We concluded that DYCI-1::mCherry is very likely to remain associated to the other members of the dynein complex in the cytoplasm.

Overall, the tube assay and the FCS experiments suggest that DYCI-1::mCherry is a faithful reporter of dynein cortical pulling during mitosis, and probably also of dynein's other roles in zygotic division.

## 2 CHARACTERIZING DYCI-1::MCHERRY DYNAMICS.

### 1.2.1 USING TWO INDEPENDENT APPROACHES FOR COUNTING MOLECULES BOUND TO THE MICROTUBULE PLUS-ENDS.

To estimate the number of DYCI-1::mCherry molecules at the growing ends of the astral microtubules (plus-ends), we used FCS. Because of the weak fluorescence intensity of the DYCI-1::mCherry spots, it can be challenging to detect peaks that correspond to a dynein spot crossing the FCS volume (see for example Fig. 3c). We therefore needed a way to determine the spot positions independently of the DYCI-1::mCherry spots. To do so, we used the doubly labelled DYCI-1::mCherry and EBP-2::GFP strain, where EBP-2 revealed the spot positions, and we assumed similar dynein spot binding kinetics between this strain and TH163, which carries the randomly integrated DYCI-1::mCherry construct. Parameters obtained for association kinetics (see 1.1.5, and Fig. S2c) were plugged into equation S2 along with the previously measured cytoplasmic concentration of 32 ± 11 molecules in the FCS focal volume ($N$ = 8 embryos, 38 spots). This resulted in an estimate of the number of particles inside a spot in a doubly labelled strain of 66 ± 5 dyneins ($N$ = 8 embryos, 43 spots).

To strengthen our results, we tried to reproduce them using a different approach, this time based on comparing spot intensities to a reference based on (Shivaraju et al., 2012). We again used the PAR-6::mCherry strain as a reference and calibrated intensity to it in order to be able to convert this into a particle count. We did so by comparing the background intensities in the images of this strain (Fig. S6d) to the cytoplasmic concentrations measured by FCS as described above ($N$ = 8 embryos, 16 spots; see 1.1.4). We then imaged PAR-6::mCherry and the strain In the strain carrying randomly integrated DYCI-1::mCherry (TH163) in identical conditions (Fig. S6a-c), fitting the dynein spot intensity profiles by a Gaussian distribution with an additional constant $b$ representing the background:

$$I = (A-b)e^{-\frac{(x-\bar{x})^2}{2\sigma^2}} + b \tag{S3}$$

Here, $\bar{x}$ is the spot's position in the intensity profile, and $\sigma^2$ is its width (e.g. Fig S6e), and $A$ the amplitude of the peak. We repeated this experiment and plotted the corresponding histogram (Fig. S6f). The resulting average was 50 ± 13 DYCI-1::mCherry particles per spot ($N$ = 6 embryos, 20 spots) in TH163, consistent with our previous estimate.

We also wished to compare this dynein-per-spot count to the EBP-2::GFP counts. To measure this, we used the doubly labelled EBP2::GFP DYCI-1::mCherry strain. We calculated the cytoplasmic concentration by FCS as explained above, which resulted in 82 ± 38 EBP-2::GFP molecules in the 0.3 fl FCS volume. We also measured the cytoplasmic intensity and used the results to calibrate the relationship between intensity and EBP-2::GFP counts. We analysed EBP-2::GFP spots at the microtubule plus–ends crossing the FCS focal volume (Fig. 3c). Assuming that the basal levels corresponded to the previously measured background, we estimated the prorata number of particles at the peak. We found 185 ± 85 EBP-2::GFP per spot ($N$ = 8 embryos, 38 spots). This is consistent with the previously calculated count for DYCI-1::mCherry, which is putatively bound to EBP.

### 1.2.2 COLOCALIZING DYCI-1::MCHERRY AND THE LATTICE OR THE PLUS-ENDS OF ASTRAL MICROTUBULES.

Directed motion of dynein spots towards the periphery (see Fig. 2a-e and the main text) is best represented by a flow model (Fig. 2f) compatible with active transport (e.g. along the microtubule lattice by a kinesin molecular motor). On the other hand, their motion is strongly reminiscent of EB protein comets which appear to track plus-ends (Akhmanova and Steinmetz, 2015), but simply accumulate at the tip of microtubules, a mechanism that does not transport EB molecules (Bieling et al., 2007). To gain an initial insight and orient our investigations between active transport or passive microtubule plus-end accumulation, we set out to confirm that dynein is accumulated at the microtubule plus-ends. To do so, we crossed the strain carrying randomly integrated DYCI-1::mCherry with one carrying EBP-2::GFP. Indeed, EBP-1, -2, and -3 are the three nematode EB protein homologs. We found that most of the DYCI-1::mCherry spots colocalized with EBP-2::GFP in the doubly labelled strain (Fig. 3a and Movie S10).

To be more quantitative, we tracked spots on both channels, and set as colocalized spots that are closer than 4 pixels at each time. We studied only those tracks longer than 6 time-points and which displayed directed motion, although we reduced the threshold track length to 3 points for cortical colocalizations. In all cases, we wondered whether the high density of DYCI-1::mCherry spots might cause artefactual colocalization. Therefore, for each colocalization experiment we compared the results with the colocalization of a synthetic set of spots of identical count produced using the same method and which were randomly distributed in the image (Jaqaman et al., 2011). In no experiment did we find a significant colocalization with the fabricated images (see for example Fig. S10b and 3b).

Dynein spots colocalized strongly in the case of DYCI-1::mCherry and EBP-2::GFP (Fig. 3b). Similarly, these two proteins crossed the FCS focal volume simultaneously, and their detected peaks coincided ($N$ = 8 embryos, 43 spots; Fig. 3c). We concluded that most of the dynein spots localize at the microtubule plus-ends.

However, we noticed that about 20% of the dynein spots did not colocalize at all. This could either be due to the limits of detection (because of the cytoplasmic fraction of EBP-2::GFP), or because these spots are along the microtubule lattice. To exclude this latter possibility, we crossed strains carrying the DYCI-1::mCherry and α-tubulin::YFP transgenes. As expected, DYCI-1 spots strongly colocalized with astral microtubules (Movie S11 and Fig. S10). Interestingly, the fraction of colocalized dynein spots was almost equal with the microtubules and with the microtubule plus-ends, suggesting that the vast majority of detectable DYCI-1::mCherry spots colocalize with the plus-ends.

### 1.2.3 CHARACTERIZING DYNEIN UNBINDING DYNAMICS AT THE MICROTUBULE PLUS-ENDS BY MODULATING MICROTUBULE GROWTH RATES.

We set out to characterize the detachment of dynein from spots at microtubule plus-ends, and how this compares to the known EBP-2$^{EB}$ dynamics (Akhmanova and Steinmetz, 2015; Bieling et al., 2007). In other words, how does the length of the "comet tail" of plus-end dynein accumulation vary when microtubule growth rates change? We plotted an intensity profile across dynein spots and used an exponential fit to measure the characteristic length of the comet tail. To change microtubule growth rates, we used hypomorphic RNAi of *klp-7*$^{MCAK}$ and *clip-1*$^{CLIP170}$, which are genes that alter microtubule dynamics (Fig. S2a) (Srayko et al., 2005). Importantly, upon penetrant RNAi, we did not see dynein depletion from the plus-ends (Fig. 4b, S5l). When we investigated dynein detachment dynamics, we measured "comet tail" length under four conditions (ordered by decreasing microtubule growth rates): *clip-1(RNAi)* during 48 h ($N$ = 5 embryos, 3960 spots, 58 profiles); *clip-1(RNAi)* during 24 h ($N$ = 6 embryos, 2781 spots, 30 profiles); untreated embryos ($N$ = 8 embryos, 3000 spots, 30 profiles); and *klp-7(RNAi)* during 24 h ($N$ = 6 embryos, 1067 spots, 30 profiles) (Fig. S2a). We observed a linear correlation between the comet tail lengths and the microtubule growth rates (Fig. 3e). We concluded that EBP-2$^{EB}$ and DYCI-1 display the same unbinding dynamics at the microtubule plus-ends.

### 1.2.4 MEASUREMENT OF CORTICAL PULLING FORCES USING ANAPHASE SPINDLE OSCILLATIONS.

Since dynein is needed to generate forces that position the spindle, as reflected by anaphase oscillations (Gonczy et al., 1999; Nguyen-Ngoc, 2007; Pecreaux et al., 2006), a defect in dynein targeting to the cell cortex should strongly reduce these oscillations, as in *gpr-1/2(RNAi)* and *dli-1(RNAi)* (Pecreaux et al., 2006). In fact, we demonstrated that only a partial reduction in the number of active force generators is needed to suppress oscillation. Despite the suggestion that a secondary dynein-independent mechanism can generate pulling forces at the cortex (Schmidt et al., 2005), this would not fully compensate for the lack of functional dynein at the cortex (Pecreaux et al., 2006) (see main text for further discussion). It therefore appears reasonable to use oscillations to investigate dynein dynamics at the cortex.

**Investigating the role of EBP-2**

EBP-2 contributes to the accumulation of dynein at the microtubule plus-ends, but surprisingly *epb-2(RNAi)* was reported to have no early embryonic phenotype (Kamath et al., 2003; Sonnichsen et al., 2005). We tracked the γTUB::GFP-labelled posterior centrosome and analysed its oscillations upon *ebp-2(RNAi)* or in the *ebp-2(gk756)* null mutant. In both cases, oscillations were reduced but still there (Fig. S4c). Because *C. elegans* has three EB protein homologs, none having an early embryonic phenotype (Kamath et al., 2003; Sonnichsen et al., 2005), we tested a putative redundancy. We treated *ebp-2(gk756)* null mutants with *ebp-1(RNAi)* or *ebp-1/-3(RNAi)* using a single transcript, but found no further reductions in oscillation amplitudes (Fig. S4c). Consistently, Schmidt and colleagues observed that in the absence of EBP-1/3, dynein is still localized at the microtubule plus-ends (Schmidt et al., 2017). We concluded that EBP-2, but not EBP-1 or EBP-3, is likely to be involved in targeting dynein to the cortex by contributing to the accumulation of dynein at the astral microtubule plus-ends. Furthermore, because some pulling force remained even without EBP-2, we suggest that there also exists a second mechanism which is partly redundant.

**Investigating the role of kinesins**

The EBP-2-dependent mechanism that causes dynein to accumulate at the microtubule plus-ends does not fully account for dynein targeting to the cell cortex. In budding yeast, the kinesin Kip2p (which has no known homolog in *C. elegans*) is involved in a mechanism for transporting dynein along microtubules (Markus and Lee, 2011; Markus et al., 2009; Roberts et al., 2014). And more generally, we know that kinesins transport dynein towards the cell periphery in other contexts (Hancock, 2014). We therefore tested the cortical force generation in γTUB::GFP embryos after RNAi silencing of each gene that encodes a kinesin motor domain. These later were the member of the corresponding InterPro family (Mitchell et al., 2015). We excluded *vab-8* as it is not expressed in the embryo (Wolf et al., 1998). We observed decreased oscillation amplitudes in just klp-13, -18, -19, and -20, although the decreases were not significant (Fig. S4). This is consistent with the results of phenotypic screens performed in nematodes which suggest that no kinesin RNAi cancels out anaphase oscillation (Kamath et al., 2003; Sonnichsen et al., 2005). After excluding KLP-18 since this was previously shown to have no role in mitosis (Segbert et al., 2003), we tested the others kinesins for which we saw phenotypes, and dynein plus-end accumulation was not decreased (Fig. S5l). We concluded that kinesin transport of dynein is not likely to be involved as a secondary targeting mechanism for making EB proteins at least partly expendable.

### 1.2.5 ANALYSIS OF CORTICAL PULLING FORCE IMBALANCES USING THE TUG-OF-WAR MODEL

As a supplement to the analysis of dynein dynamics at the cortex, we next used oscillations to study the mechanism that causes the cortical pulling force imbalance.

**A force imbalance based on the differential regulation of intrinsic force generator properties**

We set out to compare the oscillations of both centrosomes during anaphase. Likely because the spindle is weakened in early anaphase (Maton et al., 2015; Mercat et al., 2016), the coupling of the oscillation of both poles is reduced and thus some differences can be observed between the anterior and posterior sides. We only explored the amplitude and frequency predictions. Modelling these oscillations, we proposed that the amplitude positively depends mostly on the so-called negative damping $\Xi$, while the frequency scales inversely according to the square root of inertia, $I$ (Pecreaux et al., 2006):

$$\Xi = 2N\left(\frac{\bar{f}}{f_c}\bar{p}\left((1-\bar{p}) - \frac{f_c}{\bar{f}}\right)\right)f'$$

$$I = 2N\left(\frac{\bar{f}}{f_c}\bar{p}(1-\bar{p})\right)f'\bar{\tau}$$

$N$ is the number of force generators and would be different in the cortex halves upon asymmetrical total counts (possibility 1 in the main text). $\bar{p}$ is the fraction of force generators pulling at a given time and $\bar{\tau}$ a time lag. These two parameters only depend on force generator dynamics, and their asymmetry points to possibilities 2 and 3. Finally, $\bar{f}$ is the generator stall force, which corresponds roughly to the force per generator because they act at very low speed, thus close to stall force. This is due to the force-velocity relationship, which states that the pulling force of a generator is inversely proportional to its velocity, with a slope $f'$. $f_c$ is the detachment rate's sensitivity to force, and the larger this is, the less the force generator will detach when the pulling force increases. These last three parameters model the intrinsic properties of the force generator. Qualitatively, when these parameters change, we can expect variations in the frequency and amplitude of oscillations (see Table S1). The direction indicated corresponds to a higher force on the posterior side, which will ensure spindle displacement (Pecreaux et al., 2006):

| Parameter (posterior vs anterior) | Stall force $\bar{f}$, increased on post. | Force-velocity slope $f'$, increased on post. | Detachment force sensitivity $f_c$, increased on post. |
|---|---|---|---|
| Frequency | Decreased | Decreased | Increased |
| Amplitude | Increased | Increased | Decreased |

**Table S1**: Tug-of-war model predictions for increases in posterior force due to an asymmetry in intrinsic force generator (dynein) properties.

Overall, we concluded that if the force imbalance is caused by an asymmetry in intrinsic generator properties, amplitude and frequency should vary (between the anterior and posterior sides) in the opposite direction.

**Asymmetry in centrosomal anaphase oscillation characteristics**

We investigated oscillations in untreated embryos (same as Fig. S4), measuring peak-to-peak oscillation amplitudes to be 6.0 ± 1.3 µm (mean ± SD) on the posterior, and 3.1 ± 1.1 µm on the anterior ($N$ = 20). The frequencies were 47.3 ± 7.0 mHz and 44.8 ± 6.0 mHz, respectively. Although these frequencies are not significantly different, likely due to the remaining coupling of posterior and anterior centrosomal oscillations by the spindle, this does not seem to be consistent with the above prediction, suggesting that force imbalance is probably not caused by an asymmetry in the intrinsic force generator properties. Supporting this view, asymmetries in stall forces and force-velocity relationships do not result in asymmetrical active generator counts, in disagreement with previously published results (Grill et al., 2003). An asymmetry in force sensitivity would result in generators pulling for a longer time on the posterior side, thus longer residency times, which contradicts our results at the cortex (main text, Fig. 5f). Overall, from oscillation point of view, it is unlikely that the force imbalance relies on an asymmetry of intrinsic force generator properties.

**A force imbalance based on the presence of more force generators on the posterior side (total number asymmetry).**

In this scenario, we expect to have larger oscillations and forces on the posterior side, but also more inertia, thus smaller frequencies. This can be seen by considering that N and $f'$ play similar roles in the equations. In conclusion, since in our experiments the frequencies and amplitudes positively correlate between the posterior and anterior sides, the asymmetry is probably based on dynamics, as suggested in possibilities 2 and 3.

**A force imbalance based on attachment/detachment dynamics (on- and off-rates).**

The dynein detachment rate at the stall force, also called the off-rate, is modelled by $\overline{k_{off}}$, which is in turn part of the calculation of average probability $\overline{p} = k_{on}/(k_{on} + \overline{k_{off}})$ and that of the time lag $\overline{\tau} = 1/(k_{on} + \overline{k_{off}})$. The higher forces on the posterior side, which contribute to the posterior displacement of the spindle, require a lower detachment rate for that side. Alternatively, if the attachment rate, or the $k_{on}$ (on-rate), is what accounts for the asymmetry, it should be higher on the posterior side to account for the force imbalance. Processivity (the inverse of the off-rate) is likely to control mitotic progression throughout anaphase (Pecreaux et al., 2006). Therefore, an asymmetry in the detachment rate would result in a heterochrony in oscillations, since our model is linearized. In contrast, in addition to a heterochrony, an asymmetry in attachment rates would result in a higher frequency on the posterior side, assuming that processivity controls progression in the same way for both embryo halves. This is consistent with our measurements. In fact, the small difference between frequencies combined with the limitations of the model led us to conclude that both attachment and detachment rate asymmetries could account for force imbalance. Direct measurements of dynein processivity (Fig. 5f) allowed us to decide in favour of an on-rate asymmetry.

### 1.2.6 MEASURING DYNEIN RESIDENCY TIME AT THE CORTEX.

To measure the residency time of DYCI-1::mCherry at the cortex, we imaged the embryo during metaphase at the cortex plane, moving the focus down until the embryo shape appeared diffuse. Next, the focus was moved up less than one micron to recover the embryo shape, and set to this plane for imaging. The detected spots were tracked with u-track software and divided in two populations (classification): direct and diffusive. We then computed the histogram of track durations for each population and averaged them over the embryos, only taking into account tracks longer than 3 frames (600 ms). We estimated the residency time to be the characteristic time $\mu = 1/\lambda$ by fitting the averaged histograms into an exponential distribution (Fig. 2c):

$$1/\lambda * exp(-1/\lambda * t)$$

where *t* represents time.

### 1.2.7 3D DIFFUSION ESTIMATION OF DYNEIN ARRIVING AT THE CORTEX.

Using (Von Smoluchowski, 1917), we can estimate the rate of dynein reaching the cortex by 3D diffusion in an embryo modelled by a half-spheroid for each centrosome. We have assumed that return of dynein to the cell centre is done at a non-limiting rate, the equation is $r_{th} = 2\pi DRc_0$. The diffusion coefficient is $D = 2.6$ μm$^2$/s (see above). R = 17.0 nm is the hydrodynamic radius of the dynein dimer (Trokter et al., 2012), and $c_0 = 3.2 \times 10^{18}$ dynein/m$^3$ is the concentration as estimated above. The resulting theoretical rate $r_{th}$ = 30 dyneins reaching the cortex every second is in the correct order of magnitude, although at the low end of the 20 to 200 expected from the 10 to 100 active force generators per half cortex (Grill et al., 2003) that stay about 0.5 s (this work) and (Pecreaux et al., 2006), which would result in a 4 folds larger estimate.

### 1.2.8 STATISTICAL ANALYSIS OF THE DYNEIN TRACKS ASYMMETRY AT THE CELL CORTEX.

When investigating the dynein tracks at the cortex (main text, last results section), to assess whether their distribution is homogeneous or biased toward the posterior, we considered the two tip regions (1 and 4 in Fig. 5e) and applied a Bayesian approach to analyze the counts of tracks. We used raw track counts and thus ignored the variations of the region area between embryos since by design, region 1 and 4 have exactly the same area within each embryo. Since we only focus on the asymmetry of the distribution, we have not normalized by the duration of individual embryos recordings. We modeled the occurrence of a dynein track at the posterior side as coming from a binomial distribution with an unknown probability $p$ which we estimated from the dataset. To do so, we combined the likelihood estimators of $p$ obtained over all the embryos from the same condition in order to obtain a maximum likelihood estimation of $p$, as well as a confidence interval. Probability $p$ equal or smaller to 0.5 corresponds to dynein tracks distribution homogeneous or enriched on anterior side. To test the likelihood of such a null hypothesis, we integrated the density function over this range to find the p-values reported below (Table S2) and depicted by stars in Fig. 5. Overall, they suggest a higher count of tracks of dynein at the posterior tip compared to anterior one.

| Tracks with | Control (N=7) | *Gpr-1/2(RNAi)* (N=11) |
|---|---|---|
| Diffusive-like motion | $1.1 \times 10^{-11}$ | $4.5 \times 10^{-19}$ |
| Directed motion | $6.3 \times 10^{-12}$ | $2.2 \times 10^{-18}$ |

**Table S2**: p-values testing the probability that dynein tracks distributed homogeneously or were enriched anteriorly at the cortex of nematode embryo. This null hypothesis is rejected in all cases, suggesting that dynein tracks are more likely at the posterior side of the cell.

These p-values not only depend on the "level of asymmetry", but also on the number of embryos per condition, the number of tracks recorded in each of those, as well as on biological variability. Therefore, while this approach is instrumental to compare track counts between region within the same condition, it is not possible to compare conditions (control and *gpr-1/2(RNAi)* for example) by simply comparing these p-values.

## 3 IMAGE PROCESSING PIPELINE FOR IDENTIFYING DYNEIN DYNAMICS

### 1.3.1 PREPROCESSING THE IMAGES

Since dynein spots are very weak, we denoised the images to filter out the contributions from the cytoplasmic fraction and to increase the signal-to-noise ratio. Noise reduction in this manner usually relies on the assumptions that the noise is non-correlated in space and time, and that it follows a Gaussian or Poisson distribution. We reasoned that since there is a threshold of dyneins per spot, the spots under the threshold might contribute to the background and furthermore, because of their large size, they can create space and time correlations of the background "noise". We therefore opted to use CANDLE filtering/denoising (Coupe et al., 2012) (Fig. S7a,b). For this, we used the following parameters, which allowed for a proper view of the fine structures and enabled us to distinguish close individual spots: a smoothing parameter beta of 0.05; a patch radius of 1 (3x3x3 voxel); and a search volume radius of 3. We did fast processing of the dark background since normal processing did not improve the images. The processed images were then submitted to spot enhancement (Sage et al., 2005) using a Laplacian of Gaussian filter with standard deviation $\sigma = 1.25$ (Fig. S7c).

### 1.3.2 AUTOMATED TRACKING OF DYCI-1::MCHERRY FLUORESCENT SPOTS.

Because multiple tracks are present and could cross each other, we needed an algorithm with robust linking. We opted for u-track (Jaqaman et al., 2008) using the parameters shown below. We validated these parameters by analysing fabricated images having known dynamics (see 1.3.4), finding good colocalization between the simulation-stipulated and recovered tracks (Fig. S8f). We obtained the track densities by dividing the raw count of tracks by the duration of the acquisition and the area of the embryo. When observing the LSP, we excluded the spindle, which we obtained by a semi-supervised segmentation. Because we were conservative in the u-track algorithm parameters, it is possible that some long tracks were broken into pieces. Furthermore, microtubule plus-ends are only briefly in the focal plane when viewing the LSP.

| Detection | |
|---|---|
| Gaussian standard deviation | Iterate to estimate Gaussian standard deviation. Maximum number of iterations 10 |
| Rolling window time-averaging | 3 |
| Iterative Gaussian mixture-model fitting | No |
| Tracking | |
| Maximum gap to close | Cytosol (8), cortex (3) |
| Merge split | 0 |
| Minimum length of track segments from first step | Cytosol (3), cortex (6) |
| Cost function frame-to frame linking | |
| Flag for linear motion | 1 |
| Allow instantaneous direction reversal | Cytosol (0), cortex (1) |
| Search radius lower limit | 2 |
| Search radius upper limit | 5 |
| Standard deviation multiplication factor | 1 |
| Nearest neighbor distance calculation | 1 |
| Number of frames for nearest neighbor distance calculation | 9 |
| Cost function close gaps | |
| Flag for linear motion | 1 |
| Search radius lower limit | 2 |
| Search radius upper limit | 5 |
| Standard deviation multiplication factor | 3 |
| Nearest neighbor distance calculation | |
| Number of frames for nearest neighbor distance calculation | 9 |
| Penalty for increasing gap length | 1.5 |
| Maximum angle between linear track segments | 30 |

**Supplementary Table S3**: The u-track software parameters used for tracking.

### 1.3.3 CLASSIFICATION OF THE TRACKS

To characterize the dynamics of the spots, we classified the tracks according to three features: (1) the directionality; (2) the moving sense (centrifugal/centripetal); and (3) the motion model (flow/diffusion; see Fig. S9 for a typical velocity map).

1. Classification of tracks according to their asymmetry.

Visual inspection of tracks (Fig. 2a,b) suggested that some might be directed. We therefore divided them between anisotropic (directed) or isotropic (diffusive-like), based on their asymmetrical trajectories and as per the method proposed in (Huet et al., 2006). Tracks shorter than 3 frames (at the cortex) or 5 frames (in the LSP) were ignored. We chose an alpha parameter (the classification threshold) of 0.1 (90$^{th}$ percentile).

2. Classification of linear trajectories according to their moving sense.

We expected different molecular mechanisms in the directed motion of tracks causing motion in different ways, either towards the centre or towards the periphery. To classify these trajectories according to their motion sense, we first segmented the embryo contour using a supervised segmentation. For each track, we computed the Euclidean distance between each track point and the embryo contour curve, which forms a vector listing the distances to cortex that is of the same length as the track itself. The differences between adjacent elements in the vector were used to reveal whether a step in the track brought the spot closer to or farther from the cortex. We then computed the probability $\theta$ of moving towards the cortex as the ratio between the number of steps that contributed to get the spot closer to the cortex and the total number of steps in the track. Tracks with results above 0.7 were declared to be moving towards the cell cortex. Similar ratios and thresholds were used, *mutatis mutandis*, to classify tracks moving towards the centrosome. When doing so, we threw out the few tracks that showed no clear direction. Fig. 2a,b,e shows good examples of the results.

3. Classification according to various motion models.

To characterize the motion of the spots in each of the directionality and sense classes, we decided to estimate the probability of the tracks in a given class to display either a classic flowing or diffuse motion. We used the Bayesian classifier as implemented by Monnier and co-authors (Monnier et al., 2012), referred to hereafter as "BCA." We tested three alternative models: normal diffusion (D); flow (V); and flow mixed with diffusion (DV). These were tested in the LSP. At the cortex, we used diffusion (D); anomalous-diffusion (AD); and confined diffusion (CD) models (Saxton, 1994). To do so, we computed the conditional probability of each model knowing the data, using the mean square displacement (MSD) with a given time lag (Fig. S12). We repeated the probability calculations for time lags ranging from 6 to 16 frames in the LSP, and from 3 to 16 frames at the cortex, then averaged the results. Finally, in order to estimate the speed using flow and mixed flow-diffusion models, we retained the time lag that resulted in the highest conditional probability, and retrieved the speed from the parameters of the fitted model, (Monnier et al., 2012). Typical results are reproduced in Fig. 2f,g.

### 1.3.4 ANALYSING SIMULATED MICROSCOPY IMAGES TO VALIDATE THE IMAGE PROCESSING AND DATA ANALYSIS PIPELINE.

To ensure that our image processing pipeline faithfully describes spots dynamics, we created synthetic fluorescent images which mimicked our experimental data (Costantino et al., 2005) (Fig. S8a). Going further, we simulated the stochastic trajectories of particles. These underwent either a pure diffusion as per $x_{i,j}(t+1) = x_{i,j}(t) + \xi\sqrt{2Dt}$ (Movie S8), or a diffusion with an added flow, $x_{i,j}(t) = x_{i,j}(t) + \xi\sqrt{2Dt} + v_{i,j}t$ (Movie S9). Here, $x_{i,j}(t)$ represents the coordinates in two dimensions at time $t$; $\xi$ is a random number; $D$ is the diffusion coefficient; and $v_{i,j}$ is the flow speed. The track durations (lengths) were sampled from an exponential distribution. The intensity was set to be similar to the experimental one, and encoded by the Qyield quantum yield parameter. We plotted the instantaneous positions and applied a Gaussian filter to mimic the effect of the point-spread function in fluorescence microscopy. We then added two types of noise. First, we imitated the background noise by adding to each pixel a sampling of a Gaussian distribution normalized to $\varepsilon$ as per the formula $A_{noisy} = A + \epsilon M$ and corresponding to a signal-to-noise ratio of max($A$)/$\varepsilon$. Secondly, we mimicked the fluorescent background by superimposing a large number of fast-diffusing particles on the noisy image. This simulation provided a realistic scenario for testing the image processing and data analysis pipeline, and details of the parameters used are listed in Table S4.

| | |
|---|---|
| Image size | 250 x 250 pixels |
| Duration | 100 frames (20 s) |
| Density of particles (tracks) | (0.15 particles/µm$^2$) |
| Density of fast-diffusing background particles | 900 particles/µm$^2$ |
| Qyield | 0.42 (mCherry) |
| Pixel size | 0.130 nm |
| Sampling rate | 0.2 s |
| PSF type | Gaussian |
| PSF size | 0.3 µm |
| Bits | 12 |
| Diffusion coefficient | 0.002 µm$^2$/s |
| Flow speed (x and y coordinate) | 0.4 µm/s |
| Background noise standard deviation, σ | 0.1, 0.3, 0.5, 0.7 |
| Mean track lifetime (mean of exponential distribution) | 10 |
| Diffusion coefficient of the background particles | 2.1 µm$^2$/s |
| Mean track lifetime (mean of exponential distribution) for background particles | 2 s |

**Table S4**: Parameters of simulated dynein dynamics images.

We processed these fabricated images as we had the real ones (Fig. S8b,c), then analysed them. A comparison of the two data sets recovered by our analysis pipeline and from the simulation (ground-truth results) suggests that the pipeline performed well in the signal-to-noise ratio range, where we were experimentally (Fig. S8d-g and main text).

## 4 REFERENCES

Akhmanova, A., and Steinmetz, M.O. (2015). Control of microtubule organization and dynamics: two ends in the limelight. Nat Rev Mol Cell Biol *16*, 711-726.

Bieling, P., Laan, L., Schek, H., Munteanu, E.L., Sandblad, L., Dogterom, M., Brunner, D., and Surrey, T. (2007). Reconstitution of a microtubule plus-end tracking system in vitro. Nature *450*, 1100-1105.

Boulin, T., and Bessereau, J.L. (2007). Mos1-mediated insertional mutagenesis in Caenorhabditis elegans. Nat Protoc *2*, 1276-1287.

Costantino, S., Comeau, J.W., Kolin, D.L., and Wiseman, P.W. (2005). Accuracy and dynamic range of spatial image correlation and cross-correlation spectroscopy. Biophys J *89*, 1251-1260.

Coupe, P., Munz, M., Manjon, J.V., Ruthazer, E.S., and Collins, D.L. (2012). A CANDLE for a deeper in vivo insight. Med Image Anal *16*, 849-864.

Dernyi, I., Jlicher, F., and Prost, J. (2002). Formation and Interaction of Membrane Tubes. Physical Review Letters *88*, 238101.

Garrard, S.M., Capaldo, C.T., Gao, L., Rosen, M.K., Macara, I.G., and Tomchick, D.R. (2003). Structure of Cdc42 in a complex with the GTPase-binding domain of the cell polarity protein, Par6. EMBO J *22*, 1125-1133.

Gonczy, P., Pichler, S., Kirkham, M., and Hyman, A.A. (1999). Cytoplasmic dynein is required for distinct aspects of MTOC positioning, including centrosome separation, in the one cell stage Caenorhabditis elegans embryo. J Cell Biol *147*, 135-150.

Grill, S.W., Howard, J., Schaffer, E., Stelzer, E.H., and Hyman, A.A. (2003). The distribution of active force generators controls mitotic spindle position. Science *301*, 518-521.

Hancock, W.O. (2014). Bidirectional cargo transport: moving beyond tug of war. Nat Rev Mol Cell Biol *15*, 615-628.

Hirano, Y., Yoshinaga, S., Takeya, R., Suzuki, N.N., Horiuchi, M., Kohjima, M., Sumimoto, H., and Inagaki, F. (2005). Structure of a cell polarity regulator, a complex between atypical PKC and Par6 PB1 domains. J Biol Chem *280*, 9653-9661.

Howard, J. (2001). Mechanics of motor proteins and the cytoskeleton (Sunderland, Mass.: Sinauer Associates, Publishers).

Huet, S., Karatekin, E., Tran, V.S., Fanget, I., Cribier, S., and Henry, J.P. (2006). Analysis of transient behavior in complex trajectories: application to secretory vesicle dynamics. Biophys J *91*, 3542-3559.

Jaqaman, K., Kuwata, H., Touret, N., Collins, R., Trimble, W.S., Danuser, G., and Grinstein, S. (2011). Cytoskeletal control of CD36 diffusion promotes its receptor and signaling function. Cell *146*, 593-606.

Jaqaman, K., Loerke, D., Mettlen, M., Kuwata, H., Grinstein, S., Schmid, S.L., and Danuser, G. (2008). Robust single-particle tracking in live-cell time-lapse sequences. Nat Methods *5*, 695-702.

Kamath, R.S., Fraser, A.G., Dong, Y., Poulin, G., Durbin, R., Gotta, M., Kanapin, A., Le Bot, N., Moreno, S., Sohrmann, M.*, et al.* (2003). Systematic functional analysis of the Caenorhabditis elegans genome using RNAi. Nature *421*, 231-237.

Leduc, C., Campas, O., Zeldovich, K.B., Roux, A., Jolimaitre, P., Bourel-Bonnet, L., Goud, B., Joanny, J.F., Bassereau, P., and Prost, J. (2004). Cooperative extraction of membrane nanotubes by molecular motors. Proc Natl Acad Sci U S A *101*, 17096-17101.

Markus, S.M., and Lee, W.L. (2011). Regulated offloading of cytoplasmic dynein from microtubule plus ends to the cortex. Dev Cell *20*, 639-651.

Markus, S.M., Punch, J.J., and Lee, W.L. (2009). Motor- and tail-dependent targeting of dynein to microtubule plus ends and the cell cortex. Curr Biol *19*, 196-205.

Maton, G., Edwards, F., Lacroix, B., Stefanutti, M., Laband, K., Lieury, T., Kim, T., Espeut, J., Canman, J.C., and Dumont, J. (2015). Kinetochore components are required for central spindle assembly. Nat Cell Biol *17*, 697-705.

Mercat, B., Pinson, X., Fouchard, J., Mary, H., Pastezeur, S., Alayan, Z., Gachet, Y., Tournier, S., Bouvrais, H., and Pécréaux, J. (2016). Spindle Micro-Fluctuations of Length Reveal its Dynamics Over Cell Division. Biophys J *110*, 622a.

Mi, H., Dong, Q., Muruganujan, A., Gaudet, P., Lewis, S., and Thomas, P.D. (2010). PANTHER version 7: improved phylogenetic trees, orthologs and collaboration with the Gene Ontology Consortium. Nucleic Acids Res *38*, D204-210.

Mi, H., Muruganujan, A., Casagrande, J.T., and Thomas, P.D. (2013). Large-scale gene function analysis with the PANTHER classification system. Nat Protocols *8*, 1551-1566.

Mitchell, A., Chang, H.Y., Daugherty, L., Fraser, M., Hunter, S., Lopez, R., McAnulla, C., McMenamin, C., Nuka, G., Pesseat, S.*, et al.* (2015). The InterPro protein families database: the classification resource after 15 years. Nucleic Acids Res *43*, D213-221.

Monnier, N., Guo, S.M., Mori, M., He, J., Lenart, P., and Bathe, M. (2012). Bayesian approach to MSD-based analysis of particle motion in live cells. Biophys J *103*, 616-626.

Nguyen-Ngoc, T., Afshar, K., and Gonczy, P. (2007). Coupling of cortical dynein and G alpha proteins mediates spindle positioning in Caenorhabditis elegans. Nat Cell Biol *9*, 1294-1302.

Nguyen-Ngoc, T., Afshar, K., and Gonczy, P (2007). Coupling of cortical dynein and G alpha proteins mediates spindle positioning in Caenorhabditis elegans. Nat. Cell Biol. 9 *1294-1302.*

Pecreaux, J., Roper, J.C., Kruse, K., Julicher, F., Hyman, A.A., Grill, S.W., and Howard, J. (2006). Spindle oscillations during asymmetric cell division require a threshold number of active cortical force generators. Curr Biol *16*, 2111-2122.

Redemann, S., Pecreaux, J., Goehring, N.W., Khairy, K., Stelzer, E.H., Hyman, A.A., and Howard, J. (2010). Membrane invaginations reveal cortical sites that pull on mitotic spindles in one-cell C. elegans embryos. PLoS One *5*, e12301.

Robert, V., and Bessereau, J.L. (2007). Targeted engineering of the Caenorhabditis elegans genome following Mos1-triggered chromosomal breaks. Embo J *26*, 170-183.

Roberts, A.J., Goodman, B.S., and Reck-Peterson, S.L. (2014). Reconstitution of dynein transport to the microtubule plus end by kinesin. eLife *3*, e02641.

Sage, D., Neumann, F.R., Hediger, F., Gasser, S.M., and Unser, M. (2005). Automatic tracking of individual fluorescence particles: application to the study of chromosome dynamics. IEEE Trans Image Process *14*, 1372-1383.

Sarov, M., Murray, J.I., Schanze, K., Pozniakovski, A., Niu, W., Angermann, K., Hasse, S., Rupprecht, M., Vinis, E., Tinney, M.*, et al.* (2012). A genome-scale resource for in vivo tag-based protein function exploration in C. elegans. Cell *150*, 855-866.

Sarov, M., Schneider, S., Pozniakovski, A., Roguev, A., Ernst, S., Zhang, Y., Hyman, A.A., and Stewart, A.F. (2006). A recombineering pipeline for functional genomics applied to Caenorhabditis elegans. Nat Methods *3*, 839-844.

Saxton, M.J. (1994). Single-particle tracking: models of directed transport. Biophys J *67*, 2110-2119.

Schmidt, D.J., Rose, D.J., Saxton, W.M., and Strome, S. (2005). Functional analysis of cytoplasmic dynein heavy chain in Caenorhabditis elegans with fast-acting temperature-sensitive mutations. Mol Biol Cell *16*, 1200-1212.

Schmidt, R., Akhmanova, A., and van den Heuvel, S. (2017). Normal Spindle Positioning In The Absence Of EBPs And Dynein Plus-End Tracking In C. elegans. bioRxiv, 118935.

Segbert, C., Barkus, R., Powers, J., Strome, S., Saxton, W.M., and Bossinger, O. (2003). KLP-18, a Klp2 kinesin, is required for assembly of acentrosomal meiotic spindles in Caenorhabditis elegans. Mol Biol Cell *14*, 4458-4469.

Shivaraju, M., Unruh, J.R., Slaughter, B.D., Mattingly, M., Berman, J., and Gerton, J.L. (2012). Cell-cycle-coupled structural oscillation of centromeric nucleosomes in yeast. Cell *150*, 304-316.

Sonnichsen, B., Koski, L.B., Walsh, A., Marschall, P., Neumann, B., Brehm, M., Alleaume, A.M., Artelt, J., Bettencourt, P., Cassin, E.*, et al.* (2005). Full-genome RNAi profiling of early embryogenesis in Caenorhabditis elegans. Nature *434*, 462-469.

Srayko, M., Kaya, A., Stamford, J., and Hyman, A.A. (2005). Identification and characterization of factors required for microtubule growth and nucleation in the early C. elegans embryo. Dev Cell *9*, 223-236.

Trokter, M., Mucke, N., and Surrey, T. (2012). Reconstitution of the human cytoplasmic dynein complex. Proc Natl Acad Sci U S A *109*, 20895-20900.

Von Smoluchowski, M. (1917). Versuch einer mathematischen Theorie der Koagulationskinetik kolloidaler Lösungen. Zeitschrift für physikalische Chemie *92*, 129-168.

Widengren, J., Rigler, R., and Mets, U. (1994). Triplet-state monitoring by fluorescence correlation spectroscopy. Journal of fluorescence *4*, 255-258.

Wilkins, D.K., Grimshaw, S.B., Receveur, V., Dobson, C.M., Jones, J.A., and Smith, L.J. (1999). Hydrodynamic radii of native and denatured proteins measured by pulse field gradient NMR techniques. Biochemistry *38*, 16424-16431.

Wolf, F.W., Hung, M.S., Wightman, B., Way, J., and Garriga, G. (1998). vab-8 is a key regulator of posteriorly directed migrations in C. elegans and encodes a novel protein with kinesin motor similarity. Neuron *20*, 655-666.